\documentclass[twocolumn,
amsmath,amssymb,prl,
aps,superscriptaddress,longbibliography
]{revtex4-2}

\usepackage{graphicx}
\usepackage{dcolumn}
\usepackage{bm}
\usepackage{braket}
\usepackage{mathtools}
\usepackage{bbm}
\usepackage{comment}
\usepackage{makecell}
\usepackage{afterpage}
\usepackage{siunitx}
\usepackage{nicefrac}

\newcommand\blankpage{%
	\null
	\thispagestyle{empty}%
	\addtocounter{page}{-1}%
	\newpage}

\usepackage{hyperref}
\hypersetup{
	colorlinks=true,
	linkcolor=blue,
	citecolor=blue,
	filecolor=magenta,
	urlcolor=cyan
}

\usepackage[normalem]{ulem}

\def\iu{{\rm i}}

\DeclareMathOperator{\Tr}{Tr}
\DeclareMathOperator{\Var}{Var}

\def\dif{{\rm d}}

\newtheorem{theorem}{Theorem}
\newtheorem{lemma}{Lemma}

\graphicspath{{../}{tikz/}}

\makeatletter
\DeclareFontFamily{OMX}{MnSymbolE}{}
\DeclareSymbolFont{MnLargeSymbols}{OMX}{MnSymbolE}{m}{n}
\SetSymbolFont{MnLargeSymbols}{bold}{OMX}{MnSymbolE}{b}{n}
\DeclareFontShape{OMX}{MnSymbolE}{m}{n}{
	<-6>  MnSymbolE5
	<6-7>  MnSymbolE6
	<7-8>  MnSymbolE7
	<8-9>  MnSymbolE8
	<9-10> MnSymbolE9
	<10-12> MnSymbolE10
	<12->   MnSymbolE12
}{}
\DeclareFontShape{OMX}{MnSymbolE}{b}{n}{
	<-6>  MnSymbolE-Bold5
	<6-7>  MnSymbolE-Bold6
	<7-8>  MnSymbolE-Bold7
	<8-9>  MnSymbolE-Bold8
	<9-10> MnSymbolE-Bold9
	<10-12> MnSymbolE-Bold10
	<12->   MnSymbolE-Bold12
}{}

\let\llangle\@undefined
\let\rrangle\@undefined
\DeclareMathDelimiter{\llangle}{\mathopen}%
{MnLargeSymbols}{'164}{MnLargeSymbols}{'164}
\DeclareMathDelimiter{\rrangle}{\mathclose}%
{MnLargeSymbols}{'171}{MnLargeSymbols}{'171}
\makeatother

\def\oper{{\mathcal{O}}}

\begin{document}
	
	\def\papertitle{{Shadow tomography from emergent state designs in analog quantum simulators}}
	\def\authornames{{Max McGinley and Michele Fava}}
	
	\def\oxf{{Rudolf Peierls Centre for Theoretical Physics, Clarendon Laboratory, Parks Road, Oxford OX1 3PU, United Kingdom}}
	
	\title{\papertitle}
	\author{Max McGinley}
	\affiliation{\oxf}
	\affiliation{T.C.M. Group, Cavendish Laboratory, JJ Thomson Avenue, Cambridge CB3 0HE, United Kingdom}
	\author{Michele Fava}
	\affiliation{\oxf}
	\affiliation{Philippe Meyer Institute, Physics Department, \'Ecole Normale Sup\'erieure (ENS),
		Universit\'e PSL, 24 rue Lhomond, F-75231 Paris, France}
	
	\begin{abstract}
		We introduce a method that allows one to infer many properties of a quantum state---including nonlinear functions such as R{\'e}nyi entropies---using only global control over the constituent degrees of freedom. In this protocol, the state of interest is first entangled with a set of ancillas under a fixed global unitary, before projective measurements are made. We show that when the unitary is sufficiently entangling, a universal relationship between the statistics of the measurement outcomes and properties of the state emerges, which can be connected to the recently discovered phenomeonon of emergent quantum state designs in chaotic systems. Thanks to this relationship, arbitrary observables can be reconstructed using the same number of experimental repetitions that would be required in classical shadow tomography [Huang et al.~\href{https://doi.org/10.1038/s41567-020-0932-7}{Nat.~Phys.~\textbf{16}, 1050 (2020)}]. Unlike previous approaches to shadow tomography, our protocol can be implemented using only global Hamiltonian evolution, as opposed to qubit-selective logic gates, which makes it particularly well-suited to analog quantum simulators, including ultracold atoms in optical lattices and arrays of Rydberg atoms.
	\end{abstract}
	
	\maketitle

	\textit{Introduction.---}The ability to control interactions in a many-body quantum system allows one to simulate and study other complex quantum systems of interest \cite{Manin1980,Feynman1982}. In a universal quantum computer, where logical gates can be selectively applied to a few qubits at a time, one can in principle mimic the dynamics of any Hamiltonian \cite{Lloyd1996}; however at present such devices are limited by their size and noisiness \cite{Preskill2018}. In contrast, analog quantum simulators---such as ultracold atoms in optical lattices \cite{Bloch2012,Gross2017} and arrays of Rydberg atoms \cite{Weimer2010,Barredo2016,Endres2016,Browaeys2020}---typically possess global rather than site-specific control, and as such are more tailored to synthesizing specific classes of Hamiltonian. Despite their limitations in terms of programmability, such platforms are often more scalable and less noisy than computationally universal devices, and have already been used to shed light on a wide variety of many-body quantum phenomena \cite{Greiner2002,Paredes2004,Aidelsburger2013,Schreiber2015,Choi2016,Smith2016,Bernien2017,Leseleuc2019,Ebadi2021,Jepsen2021}.

	In any such experiment, a key task is to infer the properties of some many-body state once it has been prepared. In computationally universal devices, a particularly powerful technique known as shadow tomography can be employed for this purpose \cite{Aaronson2018,Aaronson2019,Huang2020}, wherein random unitary rotations are applied before projective measurements of each qubit are made (see also \cite{Elben2018,Brydges2019,Elben2022}). Using this scheme, many properties of the state can be simultaneously estimated using a single set of experimental data, and nonlinear properties such as R{\'e}nyi entropies can also be accessed. However, measurement strategies of this kind currently involve the application of spatially inhomogeneous sequences of site-selective gates. While these operations are natural in digital quantum computation, they are not available in analog quantum simulators, wherein all degrees of freedom evolve simultaneously under some global uniform Hamiltonian. Accordingly, the set of observables that can be directly accessed therein (efficiently or otherwise) is at present much more limited.

\begin{figure}
		\centering
		\includegraphics[width=246pt]{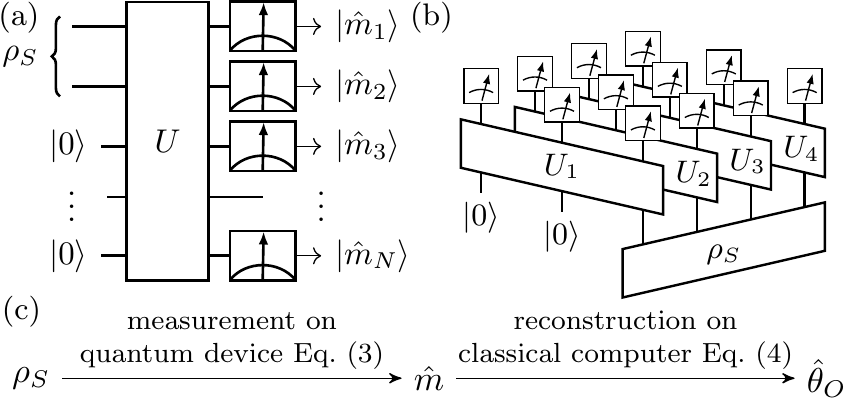}
		\caption{(a) In our protocol, the target state $\rho_S$ is evolved together with a set of ancillas under a fixed unitary $U$, before projective measurements are made in the computational basis. (b) For a many-body state, qubits can be subdivided into `blocks', each of which interact with a separate set of ancillas. (c) Schematic of the process for constructing estimators $\hat{\theta}_\oper$ of expectation values $\braket{\oper}$. 
		}
		\label{fig:Circuit}
	\end{figure}

	In this paper, we bridge this gap by introducing a new protocol that allows one to simultaneously infer many properties of a state (including R{\'e}nyi entropies, etc.) without needing to address each degree of freedom individually. Rather than applying inhomogeneous unitaries drawn randomly and compiled from few-qubit gates, we propose to apply some fixed deterministic global unitary $U$ to the system together with a set of ancillas, followed by measurements in the computational basis [see Fig.~\ref{fig:Circuit}(a)]. The unitary need not be fine-tuned, and so can be native to the system in question, making our protocol particularly well-suited to analog quantum simulators. Importantly, our scheme offers the same performance guarantees as classical shadow tomography \cite{Huang2020}, meaning that the number of measurements needed to estimate a wide range of expectation values does not grow with system size.

	We show that for generic choices of $U$, 
	a universal relationship between properties of the target state and the distribution of measurement outcomes  emerges. Specifically, the procedure becomes equivalent to making measurements of the state in bases drawn randomly from the Haar ensemble. 
	This equivalence is made precise later through our introduction of a construction called the \textit{tomographic ensemble}: a probability distribution of wavefunctions that describes the overall measurement process [Eqs.~(\ref{eq:POVMOp}, \ref{eq:Ensemble})]. For sufficiently scrambling $U$, integer moments of this ensemble agree closely with the Haar ensemble, i.e.~an approximate quantum state design (QSD) is formed \cite{Renes2004,Ambainis2007}. 
	Consequently, properties of the system density matrix can be reconstructed through appropriate post-processing of the measurement outcomes. 
	This can be achieved with moderate resources, while allowing low errors in observables' estimates ($\lesssim 1 \%$).

	The emergence of QSDs from a single global unitary (as opposed to random sequences of local gates \cite{Harrow2009}) can be related to the recently introduced concept of `deep thermalization', where QSDs
	appear in the projected ensemble of many-body quantum states \cite{Cotler2021,Choi2021,Ho2022,Claeys2022,Ippoliti2022,Lucas2022}. By adapting analytical arguments developed in that context, we rigorously establish the existence of 
	QSDs for particular representative cases. We supplement this with numerical evidence for generic choices of $U$, which allows us to benchmark the full tomography procedure, and understand the effect of symmetries. 

	\textit{Protocol.---}Our aim is to measure properties of some state of interest $\rho_S$, which is prepared at the beginning of each run of the experiment in some register $S$. For concreteness, we consider systems of qubits, although similar considerations apply to more general setups. We assume that projective measurements of all qubits can be made in some computational basis $\{\ket{m}\}$, which without loss of generality we take to be $Z_i$-diagonal, where $i$ labels the qubits and $(X_i,Y_i,Z_i)$ are Pauli operators.
	
	Projective measurements in the fixed basis $\{\ket{m}\}$ give us access to expectation values of diagonal observables, e.g.~$\braket{Z_iZ_j}$. To learn off-diagonal observables, one can apply an appropriate unitary to the system qubits before measurement. For instance, if we rotate every qubit by $e^{-\iu \pi \sum_i Y_i/4}$, then observables such as $\braket{X_iX_j}$ can be learned. However, in analog quantum simulators, where we have only global control, observables such as $\braket{X_iY_j}$ cannot be measured in this way, since different unitaries would have to be applied to qubits $i$ and $j$ separately---an operation which we assume to be unavailable. (See also Refs.~\cite{Vankirk2022,WenWei} for a discussion.)
	
	To overcome this limitation, we propose a protocol that employs a set of ancilla qubits $A$ initialized in some predetermined state, which for convenience we assume to be a pure product state $\ket{0^{\otimes N_A}}$ (this assumption is not strictly necessary). The system and ancilla qubits are jointly evolved using some fixed global unitary $U$, which is generated by a (possibly time-dependent) Hamiltonian that can be readily simulated on the platform in question. We refer to all such unitaries as \textit{native}. Finally, all qubits are measured in the computational basis $\{\ket{m}\}$. This is repeated $M$ times, resulting in a collection of $M$ bitstrings $m^{(r)} = m_1^{(r)}m_2^{(r)}\ldots m_{N}^{(r)}$, for $r = 1, \ldots M$, each of length $N = N_S + N_A$. This protocol is illustrated in Fig.~\ref{fig:Circuit}(a).
	
	Our claim is that if a native unitary $U$ is sufficiently entangling (in a sense soon to be made precise), then \textit{any} observable can be inferred from the distribution of measurement outcomes $m$, and---crucially---that the number of experimental repetitions $M$ and amount of classical computation required to estimate most observables of interest can be bounded, in the same spirit as classical shadow tomography \cite{Huang2020}. Remarkably, this is possible using just a single, fixed choice of $U$ each run (although later we will show that quantitative performance improvements can be obtained by sampling $U$ from an ensemble of native unitaries in each run).

	The above claim can be more precisely specified using the formalism of positive operator-valued measures (POVMs). Most generally, any measurement scheme on a state $\rho_S$ whose possible outcomes are indexed by $m$ can be captured by a set of positive Hermitian operators $F_m \geq 0$, known as a POVM, chosen such that the probability of obtaining outcome $m$ is $\mathbbm{P}(m|\rho_S) = \Tr[F_m \rho_S]$. The constraint that all probabilities should sum to unity implies $\sum_m F_m = \mathbb{I}_S$. For our protocol, the POVM operators are given by
	\begin{align}
		F_m =  (\mathbb{I}_S \otimes \bra{0}_A)U^\dagger \ket{m}\bra{m} U (\mathbb{I}_S \otimes \ket{0}_A).
		\label{eq:POVMOp}
	\end{align}
	We have assumed that the initial state of the ancilla is pure and the evolution is unitary. Therefore, $F_m$ are proportional to rank-1 projectors $F_m = d_S q_m \ket{\phi_m}\bra{\phi_m}$, where $q_m = \mathbbm{P}(m|\mathbb{I}_S/d_S) = \Tr[F_m]/d_S$ are the outcome probabilities for the maximally mixed state $\mathbb{I}_S/d_S$, $\ket{\phi_m}$ are normalized wavefunctions, and $d_S = 2^{N_S}$. 
	Since $q_m \geq 0$ and $\sum_m q_m = 1$, formally we can define a probability distribution over pure states on $S$, where the normalized wavefunction $\ket{\phi_m}$ occurs with probability $q_m$. We refer to this distribution, which contains complete information about the POVM, as the \textit{tomographic ensemble}.
	
	We argue that for generic choices of $U$ generated by local interactions without conservation laws, the tomographic ensemble exhibits a useful universal property, namely that it forms an approximate 
	QSD~\cite{Renes2004,Ambainis2007}. This means that for small enough integers $k$, the $k$th moments of the ensemble
	\begin{align}
		\mathcal{E}^{(k)} = \sum_m q_m \big(\ket{\phi_m}\bra{\phi_m}\big)^{\otimes k} \equiv \frac{1}{d_S} \sum_m \Tr[F_m] \tilde{F}_m^{\otimes k}
		\label{eq:Ensemble}
	\end{align}
	agree with the $k$th moments of the Haar ensemble $\mathcal{E}_{\rm Haar}^{(k)}$ up to some small error. (Here, $\tilde{F}_m = F_m/\Tr[F_m]$ are unit-trace positive operators.) Intuitively, closeness of a given ensemble to the Haar measure [as quantified by the moments \eqref{eq:Ensemble}] implies that the probability distribution covers the space of states approximately uniformly. If the dynamics $U$ respects some symmetry, then $\mathcal{E}^{(k)}$ will instead tend towards an alternative ensemble, where within each symmetry charge sector a $k$-design is formed; we discuss this case in the supplement \cite{SM}.
	
	We first provide evidence justifying the above claim, and then describe how this property can be leveraged to perform shadow tomography of target states $\rho_S$.

	\textit{Emergent quantum state designs.---}The formation of QSDs in the tomographic ensemble is reminiscent of the concept of deep thermalization. In the latter, a bipartite wavefunction $\ket{\Psi^{SA}}$ is prepared by applying a unitary $U$ to a product state, the qubits on $A$ are measured projectively, therefore producing an ensemble of states on $S$. Deep thermalization is achieved if this ensemble reproduces the Haar ensemble up to the $k$th moment for some $k>1$.
	While deep thermalization and QSDs in the tomographic ensemble are distinct concept, they bear many similarities. This connection is particularly fruitful since there are examples~\cite{Cotler2021,Ho2022,Claeys2022} where the emergence of deep thermalization can be rigorously established. We have adapted these proofs to show that the tomographic ensemble forms an (approximate) QSD when $U$ is drawn from the Haar ensemble, or is a dual-unitary circuit evolved for a time $t \geq N_S$ \cite{SM}.

	These two cases are illustrative, albeit contrived, examples where rigorous results that support our claim can be obtained. For more practical purposes, we wish to illustrate that the same occurs for generic unitaries that arise in analog quantum simulators, and for this purpose we must turn to numerical simulations.
	As figure of merit, following Ref.~\cite{Cotler2021}, we employ the trace distance $\Delta^{(k)} \coloneqq \frac{1}{2}\|\mathcal{E}^{(k)} - \mathcal{E}^{(k)}_{\rm Haar}\|_1$, which quantifies how far the tomographic ensemble is from being a $k$-design ($\| C\|_1 = \Tr[\sqrt{C^\dagger C}]$ is the trace norm).
	We study dynamics generated by Hamiltonians of the form $H(t) = \sum_j X_j X_{j+1} + h^x(t) X_j + h^y(t) Y_j { + h^z(t) Z_j}$, which approximates the native dynamics of Rydberg atom quantum simulators \cite{Bernien2017}; here $X_j, Y_j, Z_j$ are Pauli matrices for qubit $j$. In certain parameter regimes, this model is known to give rise to fast scrambling of information \cite{Banuls2011,Kim2013,Kim2014,Hosur2016}. Furthermore, when the fields $h^{x,y,z}(t)$ are time-dependent, there are no conserved quantities (including energy density), and we find that this encourages a rapid approach to $k$-design. In particular, we find that Floquet evolution works well, with $h_x = 0.8, h_z=0$, and $h_y(t)$ toggling periodically between $0.9$ for $t \in [n,n+0.5)$ and $1.8$ for $t \in [n-0.5, n)$, with $n \in \mathbbm{Z}$. In the following, the system qubits are located at the centre of a chain with open boundary conditions.

\begin{figure}
		\centering\includegraphics{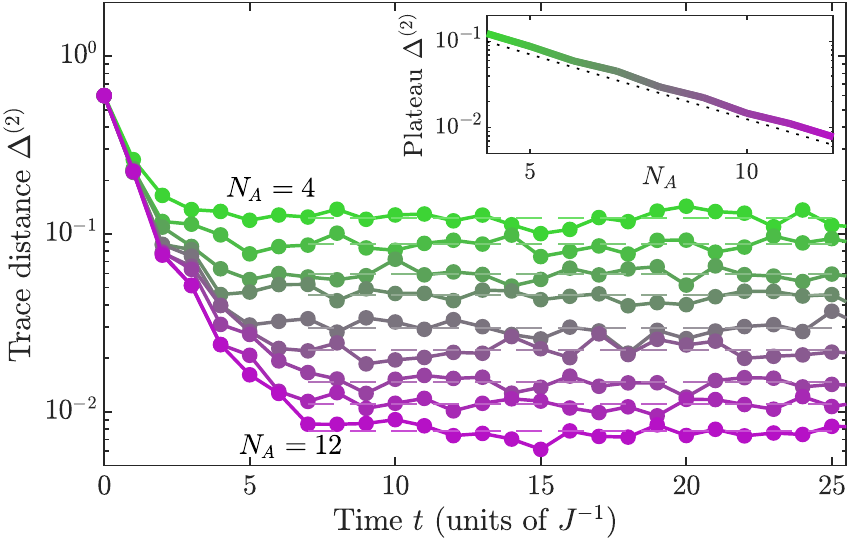}
		\caption{Trace distance between the moments of the tomographic ensemble \eqref{eq:Ensemble} and the Haar ensemble for a unitary $U = U_F^t$, with $N_S=2$ and $N_A$ increasing from 4 (green) to 12 (purple). The Floquet unitary is $U_F = e^{-\iu H_2/2}e^{-\iu H_1/2}$, with Hamiltonians $H_{1,2}$ describing the tilted field Ising model with different field values; see main text. Inset: The plateau values (dashed lines in main plot) scale approximately as $\sim 2^{-N/2}$ (dotted line).}
		\label{fig:TraceDist}
	\end{figure}

	The behaviour of the trace distance for $k=2$ as a function of time is shown in Fig.~\ref{fig:TraceDist} for various different $N_A$. We see approximately exponential decay with time, until a plateau is reached. The value of this plateau is close to the average trace distance that one obtains by replacing $\ket{\phi_m}$ with $2^{N}$ independently sampled Haar-random wavefunctions, indicating that the states making up the tomographic ensemble are effectively quasirandom. Accordingly, the plateau trace distance scales as $\sim 1/\sqrt{2^{N}}$. This behaviour is qualitatively similar behaviour to that seen in the projected ensemble of wavefunctions generated from non-energy-conserving dynamics \cite{Ippoliti2022}.
	
	\textit{Extracting properties of the state.---}Having established that the POVMs generated from our protocol generically form QSDs,
	we now describe how this property can be leveraged to efficiently learn properties of $\rho_S$. While 2-designs are known to be optimal for full reconstruction of the system density matrix \cite{Scott2006} or process tomography \cite{Emerson2005, Dankert2009}, here we describe an explicitly shadow tomographic scheme for extracting information about $\rho_S$, which in comparison keeps the sample complexity and classical computational cost bounded \cite{Aaronson2018, Aaronson2019, Huang2020}.
	
	For a fixed unitary $U$, the distribution of measurement outcomes $p_m$ depends on the state $\rho_S$ through the POVM operators \eqref{eq:POVMOp}. It will be useful to treat operators on $S$ as vectors over a $d_S^2$-dimensional space, denoted using double angled brackets $|\oper\rrangle$ and equipped with the inner product $\llangle \oper|\oper'\rrangle = \Tr[\oper^\dagger \oper']$. Similarly, the outcome distribution can be written as a $2^N$-dimensional vector $|p) = \sum_m p_m |m)$ where $|m)$ is an orthonormal basis for $\mathbbm{R}^{2^N}$, i.e.~$(m|m') = \delta_{m,m'}$. One can then define a completely positive linear map, which we call the POVM channel
	\begin{align}
		\mathcal{F} = \sum_m |m)\llangle F_m|.
		\label{eq:POVMChannel}
	\end{align}
	The observed experimental outcomes $\{\hat{m}^{(r)}\}$ ($r = 1, \ldots, M$) are evidently distributed according to the probability vector $|p) = \mathcal{F}|\rho_S\rrangle$.
	
	The inverse problem of learning properties of $\rho_S$ from experimental data $\{\hat{m}^{(r)}\}$ can be solved by finding a left inverse $\mathcal{G}$ satisfying $\mathcal{G} \mathcal{F} = \text{id}$. This allows us to construct an unbiased estimator $\hat{\theta}_\oper$ for any expectation value $\braket{\oper} = \Tr[\mathcal{O}\rho_S]$ according to $\hat{\theta}_\oper = M^{-1}\sum_{r=1}^M \llangle \mathcal{O}|\mathcal{G}|\hat{m}^{(r)})$
	In the spirit of shadow tomography \cite{Aaronson2018, Aaronson2019, Huang2020}, this estimator can be computed without needing to reconstruct the full density matrix $\rho_S$, which would be sample-inefficient. Such an inverse $\mathcal{G}$ only exists when $\mathcal{F}$ has full row rank: a condition known as \textit{informational completeness}, which is guaranteed when the tomographic ensemble forms a 2-design \cite{SM}.

	While $\mathcal{G}$ is non-unique in general, to minimize sample complexity we choose the inverse that minimizes the (average-case) variance $\text{Var}\, \hat{\theta}_\oper = \mathbbm{E}_{\hat{m}}[\hat{\theta}_\oper^2] - \mathbbm{E}_{\hat{m}}[\hat{\theta}_\oper]^2$, namely \cite{SM}
	\begin{align}
		\mathcal{G}^\ast &\coloneqq \mathcal{M}^{-1} \tilde{\mathcal{F}}^\dagger & \text{where }\mathcal{M} \coloneqq 
		\sum_m \Tr[F_m] |\tilde{F}_m\rrangle \llangle \tilde{F}_m|,
		\label{eq:InvConstruct}
	\end{align}
	where we defined the normalized channel $\tilde{\mathcal{F}} = \sum_m |m)\llangle \tilde{F}_m|$. The map $\mathcal{M}$ is a superoperator mapping the space of operators on $S$ to itself. It has full rank whenever $\mathcal{F}$ is informationally complete, and therefore has a unique inverse.

	At this point, recalling that the POVM operators \eqref{eq:POVMOp} are rank-1 projectors, we notice that the superoperator $\mathcal{M}$ is equivalent to the second moment of the tomographic ensemble $\mathcal{E}^{(2)}$, Eq.~\eqref{eq:Ensemble} \cite{SM}. Now, having established that QSDs
	generically appear in our protocol, we can replace $\mathcal{M}$ with its universal 2-design form $\mathcal{M} = (\text{id} + |\mathbbm{I}\rrangle\llangle \mathbbm{I}|)/(d_S+1)$, which has an inverse 
	\begin{align}
		\mathcal{M}^{-1}[\oper] = (d_S+1)\oper - \Tr[\oper]\,\mathbbm{I}.
		\label{eq:MInverse}
	\end{align}
	By using the fact that a 2-design is formed, we circumvent having to explicitly compute $\mathcal{M}^{-1}$, which keeps the classical computational cost bounded.
	
	Using $\llangle \mathcal{O}|\mathcal{G}^*|m) = \llangle \mathcal{M}^{-1}[O]| \tilde{F}_m\rrangle$, we can express the variance of the estimator \eqref{eq:InvConstruct} as
	\begin{align}
		\Var\,\hat{\theta}_\oper &= \frac{1}{M} \Tr\Bigg[ (\rho_S \otimes \mathcal{M}^{-1}[\oper]^{\otimes 2})
		\left(\sum_m F_m \otimes \tilde{F}_m^{\otimes 2} \right) \Bigg]
		\label{eq:Variance3Design}
	\end{align}
	The factor in rounded brackets we identify as the third moment, $\mathcal{E}^{(3)}$ in Eq.~\eqref{eq:Ensemble}. Therefore, if the tomographic ensemble forms a 3-design, as we expect for generic unitaries $U$, then the variance \eqref{eq:Variance3Design} will be the same as for any other POVM for which the $F_m$ form a 3-design. One such POVM arises in classical shadow tomography with random global Clifford unitaries \cite{SM}. Therefore we can conclude that our scheme can be used to estimate expectation values of $\rho_S$ using the same number of repetitions $M$ as one would need when doing ordinary classical shadow tomography. The dependence of the variance on the observable in question is well-characterized in Ref.~\cite{Huang2020}: observables with bounded spectral norm $\|\oper\|_\infty = \sqrt{\max \text{eig} \oper^\dagger \oper}$ can be efficiently estimated for any system size $N_S$. The procedure can be generalised in the same way as classical shadow tomography to estimate nonlinear observables, e.g.~R{\'e}nyi entropies \cite{SM}.

	To summarise, we have shown that the inverse map \eqref{eq:InvConstruct} can be used to construct estimators of expectation values, and that the map $\mathcal{M}^{-1}$ can be replaced by its universal form \eqref{eq:MInverse} when the tomographic ensemble forms an approximate 2-design. The deviation from the 2-design will govern the systematic error, since $|\mathbbm{E}_m\hat{\theta}_{\oper} - \braket{\oper}| \leq (d_S+1) \Delta^{(2)}\|O\|_\infty$, where $\Delta^{(2)}$ is the trace distance, while the $k=3$ moments $\mathcal{E}^{(3)}$ determine the variance via \eqref{eq:Variance3Design}. It is evidently favourable to have the tomographic ensemble as close to a 2- and 3-design as possible, which occurs for generic chaotic evolution as we saw above.

	\textit{Benchmarking the protocol.---}We now provide numerical simulations of our full protocol, including the joint evolution of the system and ancillas, the sampling of measurement outcomes, and the reconstruction of observables. We test our measurement scheme on a family of two-qubit target states $\rho_S(\alpha) = \alpha \ket{\text{EPR}}\bra{\text{EPR}} + (1-\alpha)[\ket{00}\bra{00} + \ket{11}\bra{11}]/2$, where $\ket{\text{EPR}} = (\ket{00} + \ket{11})/\sqrt{2}$. The coherence parameter $\alpha \in [0,1]$ allows us to interpolate between fully dephased ($\alpha = 0$) and pure $(\alpha = 1)$ EPR pairs. For the purpose of demonstration, the observables we choose to reconstruct are the fidelity with the EPR state $\Tr[\rho_S \ket{\text{EPR}}\bra{\text{EPR}}]$ and the purity $\Tr[\rho_S^2]$.
	
	In one set of simulations, we generate $U$ from Floquet evolution using the tilted-field Ising model as a generating Hamiltonian, as before. In a second set, we also add some randomness to $U$---that is, for each repetition $r$ we generate a distinct $U^{(r)}$ by selecting random magnetic fields.
	Then, $U^{(r)}$ is used in the joint system-ancilla evolution, and in the construction of estimators.
	This helps to bring the tomographic ensemble closer to a $2$-design, therefore further reducing systematic errors~\cite{SM}. To construct random unitaries $U^{(r)}$, for each time interval of length $\tau = { 1}$, we sample each field component $h^{x,y,z}$ independently from a normal distribution with zero mean and standard deviation $\sqrt{2}$.

\begin{figure}
		\centering
		\includegraphics{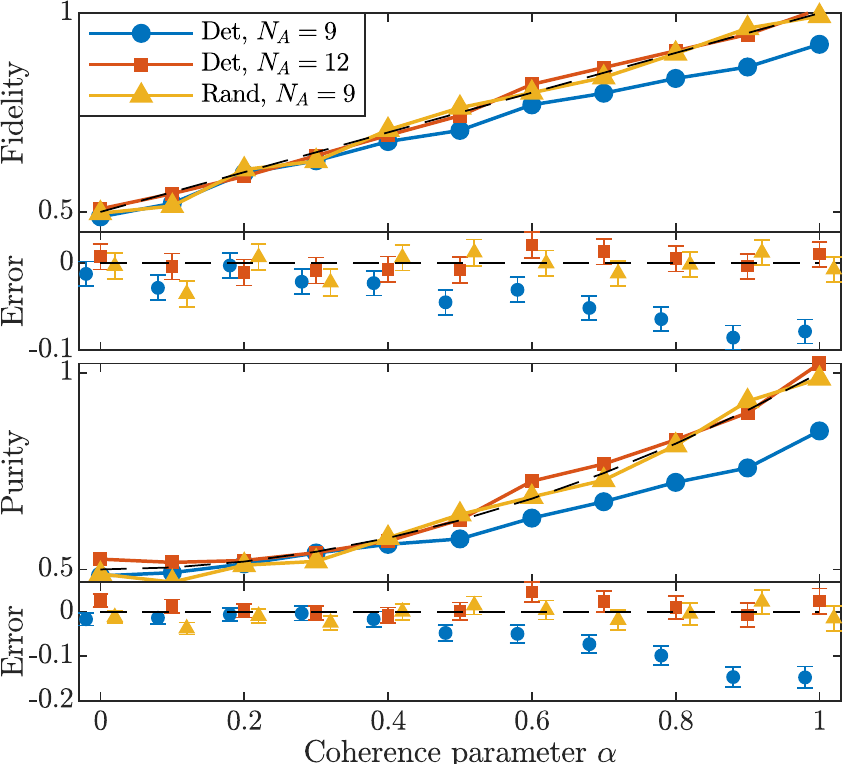}
		\caption{Estimations of the fidelity $\Tr[\rho_S\ket{\text{EPR}}\bra{\text{EPR}}]$ (top panels) and purity $\Tr[\rho_S^2]$ (bottom panels) using the deterministic protocol (Det; fixed $U$), and the semi-randomized protocol (Rand), see main text. In both cases the total evolution time is $t=10$. The target state is the EPR pair state $\ket{\text{EPR}}$ after dephasing with strength $1-\alpha$. $M = 5\times 10^3$ repetitions are used for all data points. Errors relative to the true data are shown, with data artificially shifted horizontally for readability. 
		}
		\label{fig:benchmark}
	\end{figure}

	In Fig.~\ref{fig:benchmark}, we plot estimations of the fidelity and purity for various different $\alpha$ and tomography schemes, using $M = 5\times 10^3$ repetitions each and evolving for a total time $t=10$. 
	We see closer agreement with the true fidelity as $N_A$ is increased, and when randomness is introduced.

	\textit{Classical computations.---}As in classical shadow tomography, the estimation of expectation values from experimental data requires a certain amount of classical post-processing, the complexity of which we wish to bound. Specifically, when an outcome $m$ is observed we must evaluate $\llangle \oper|\mathcal{G}_0|m)$, which requires computation of the backwards time evolution $U^\dagger \ket{m}$.
	
	When the number of system qubits $N_S$ is $O(1)$, the evolution time required to obtain an approximate QSD
	is also $O(1)$, and hence efficient matrix product state techniques can be used even for large $N_A$. For tomography of many-body states, the present strategy must be modified, since the time of evolution required to reach a QSD
	grows with $N_S$. Instead of evolving all system qubits with a single collection of ancillas, one can instead block the system into $n$ groups of $N_S/n =O(1)$ qubits, and evolve each block jointly with a separate collection of ancillas $A_{j}$ under a unitary $U_j$, where $j = 1,\ldots, n$. This scheme, illustrated in Fig.~\ref{fig:Circuit}(b), yields POVM operators $F_{m_1,\ldots, m_n} = \bigotimes_{j=1}^n F_{m_j}^{(j)}$, where each $F_{m_j}^{(j)}$ is of the form \eqref{eq:POVMOp}. The tomographic ensemble for each separate block reaches an approximate 3-design in a $O(1)$ time, allowing $F_{m_1,\ldots, m_n}$ to be evaluated efficiently using matrix product methods as before. The tradeoff is that $\mathcal{M}^{-1}$ must be replaced by a $n$-fold tensor product of \eqref{eq:MInverse}, and this will affect how the estimator variance \eqref{eq:Variance3Design} depends on the observable $\oper$. By analogy to shadow tomography with random local Pauli measurements \cite{Huang2020}, observables with support on a small number of blocks will still be accessible using a reasonable number of repetitions $M$, regardless of how big $S$ is; we prove bounds on the variance in the supplement that confirm this \cite{SM}.
	
	Note that one could in principle compute the map $\mathcal{M}^{-1}$ without using the universal 2-design form \eqref{eq:MInverse}, which would eliminate any systematic error in estimation. However, this is only feasible for a small number of ancillas $N_A$, since $2^{N_S+N_A}$ separate terms must be summed to construct $\mathcal{M}$.

	\textit{Note added.---}During completion of this work we became aware of a complementary study, to appear in the same arXiv posting, where a similar measurement scheme is presented \cite{WenWei}. The protocol introduced in that work follows the same steps as ours, where the state is first entangled with ancillas, before measurements in the computational basis are made, data from which are post-processed classically to infer properties of the state. In contrast to our proposal, no assumption is made about the formation of a QSD;
	instead the inverse map $\mathcal{M}^{-1}$ needs to be explicitly computed.

\begin{acknowledgments}
	\textit{Acknowledgements.---}MM thanks Shivaji Sondhi for helpful discussions. We are especially grateful to Sounak Biswas for insight throughout the completion of this work. We acknowledge support from UK Engineering and Physical Sciences Research Council Grant No.~EP/S020527/1.
\end{acknowledgments}

\bibliography{main}


\newpage
\afterpage{\blankpage}

\newpage
\begin{onecolumngrid}
	

	\begin{center}
		{\fontsize{11}{11}\selectfont
			\textbf{Supplemental Material for  ``\papertitle''\\[5mm]}}
		{\normalsize \authornames\\[1mm]}
		
	\end{center}
	\normalsize\

	\setcounter{equation}{0}
	\setcounter{figure}{0}
	\setcounter{table}{0}
	\setcounter{page}{1}
	
	\renewcommand{\theequation}{S\arabic{equation}}
	\renewcommand{\thefigure}{S\arabic{figure}}
	\renewcommand{\thesection}{S\arabig{section}}
	
	\suppressfloats
	
	\section{Proofs of quantum state designs in the tomographic ensemble}
	
	As mentioned in the main text, the construction of the tomographic ensemble [Eq.~\eqref{eq:POVMOp}] resembles that of the projected ensemble for a many-body state $\ket{\Psi^{SA}} = U\ket{0}$ using the same unitary $U$, as can be seen in Fig.~\ref{fig:compare_protocols}. Comparing the two cases, we see that in the region $A$, all qubits begin in the same initial state and are measured at the end of the process. The two protocols therefore differ only in the inputs and outputs of $U$ in the region $S$. (One could in principle also consider scenarios where measurement events occur throughout the dynamics, as in the study of `retrodiction' in noisy quantum dynamics \cite{Gammelmark2013}; however we leave this possibility for future work.)\\ 
	
	In this section, we make use of this resemblance to establish the existence of (approximate) quantum state designs in the tomographic ensemble, based on arguments that prove the same for the projected ensemble. We consider two cases: 1) where $U$ is a single unitary drawn at random from the Haar ensemble over $\mathrm{U}(2^N)$, and 2) where $U$ is a dual-unitary circuit containing $t \geq N_S$ timesteps.

	\subsection{Haar-random unitary}
	
	The statement we wish to prove is as follows
	\begin{theorem}
		For a unitary $U$ chosen at random from the Haar ensemble over $\mathrm{U}(2^N)$, the tomographic ensemble forms an $\epsilon$-approximate $k$-design with probability $1-\delta$ if
		\begin{align}
			N_A = \Omega\big(k N_S + \log(1/\epsilon) + \log \log (1/\delta)\big)
			\label{eq:HaarScaling}
		\end{align}
		where $\Omega(\,\cdot\,)$ contains a constant multiplicative prefactor, as well as constant offset terms.
		\label{thm:Haar}
	\end{theorem}
	The above implies that approximate quantum state designs are realised with overwhelmingly high probability in the limit of large system sizes (tending to unity double-exponentially fast in $N_A$), provided the number of ancillas scales quickly enough with the size of the system (at least as fast as $k N_S$).\\

	\textit{Proof of Theorem \ref{thm:Haar}.---}Our proof uses many of the same analytical tools as the proof of the analogous theorem for the projected ensemble given in Ref.~\cite{Cotler2021}. First, one shows that the average of $\mathcal{E}^{(k)}$ over all choices of $U$ matches the moments of the Haar distribution. Then, concentration of measure results can be used to bound the fluctuations of $\mathcal{E}^{(k)}$ away from its average. This can be used to upper bound the probability that the trace distance $\|\mathcal{E}^{(k)} - \mathcal{E}^{(k)}_{\rm Haar}\|_1$ exceeds an allowed tolerance $\epsilon$. The main difference between the two proofs will be the derivation of an upper bound of the variation of $\mathcal{E}^{(k)}$ considered as a function of $U$.\\
	
	\begin{figure}
		\centering
		\includegraphics{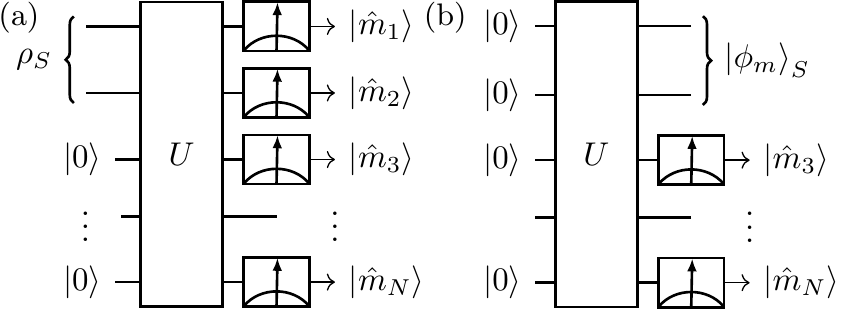}
		\caption{(a) Circuit diagram for our tomography protocol, as in Fig.~\ref{fig:Circuit}(a). (b) Protocol for generating the projected ensemble for the state $U\ket{0}_{SA}$, where $\ket{0}_{SA}$ is an extension of the product state $\ket{0}_A$ that also encompasses the system qubits. Note that the two protocols only differ in how the system qubits are prepared, and whether they are read out at the end.}
		\label{fig:compare_protocols}
	\end{figure}
	
	The calculation of the averaged moments $\mathbbm{E}_U \mathcal{E}^{(k)}$ can be achieved by first proving that the probabilities $q_m = \Tr[F_m]/d_S$ and normalized wavefunctions $\ket{\phi_m} = (\mathbbm{I}_S \otimes \bra{0}_A)U^\dagger \ket{m}/\sqrt{q_m}$ are independent random variables. To see this, note that the joint probability density for a pair $(q_m, \ket{\phi_m})$ satisfies $\dif \mathbbm{P}(q_m, \ket{\phi_m}) =\dif \mathbbm{P}(q_m, U_S \ket{\phi_m})$ by virtue of the fact that for any $U$ realizing a given pair $(q_m, \ket{\phi_m})$, there exists a unitary $U(U_S^\dagger \otimes \mathbbm{I}_A)$ occurring with the same probability (thanks to the invariance of the Haar measure), which realizes the pair $(q_m, U_S \ket{\phi_m})$. This implies that the conditional probability density $\dif \mathbbm{P}(\ket{\phi_m} | q_m)$ is invariant under unitary rotations $U_S$, and hence must be equal to the Haar measure for any $q_m$. One can therefore separate averages of $q_m$ and $\ket{\phi_m}$, and using the definition of the moments \eqref{eq:Ensemble} we have
	\begin{align}
		\mathbbm{E}_U \mathcal{E}^{(k)} \equiv \mathbbm{E}_U \sum_m q_m (\ket{\phi_m}\bra{\phi_m})^{\otimes k} = \sum_m \left(\mathbbm{E}_U q_m\right) \mathbbm{E}_{\Psi \sim \text{Haar}} \ket{\Psi}\bra{\Psi}^{\otimes k} = \mathcal{E}^{(k)}_{\rm Haar}.
		\label{eq:HaarAverageMoment}
	\end{align}
	
	We now seek to upper bound the probability of finding the $k$th moment a distance at least $\epsilon$ away from its mean. This can be done using concentration of measure results, which describe the general phenomenon where probability distributions over high-dimensional manifolds become approximately uniform (see, e.g.~Ref.~\cite{Ledoux2001}). Whereas the relevant quantity in Ref.~\cite{Cotler2021} was a functional of a Haar-random state, here the moments of the tomographic ensemble are functionals of a Haar-random unitary. The particular lemma that we will need is therefore slightly different; it is stated as Lemma 3.2 in Ref.~\cite{Low2009}:
	\begin{lemma}[L{\'e}vy's lemma]
		Given a function $f(U)$ that for any two $U_{1,2} \in \mathrm{U}(d)$ satisfies
		\begin{align}
			\frac{|f(U_1) - f(U_2)|}{\|U_1 - U_2\|_2} \leq \eta,
			\label{eq:Lipschitz}
		\end{align}
		where $\| C \|_2 \coloneqq \sqrt{\Tr[C^\dagger C]}$ is the Frobenius norm, the probability that $f$ deviates from its mean $\mathbbm{E}_U f$ by at least $\epsilon$ can be upper bounded as
		\begin{align}
			\mathbbm{P}(|f - \mathbbm{E}f| \geq \epsilon) \leq 4 \exp\left(-\frac{2d \epsilon^2}{9\pi^3 \eta^2}\right).
		\end{align}
		\label{thm:Levy}
	\end{lemma}
	The constant $\eta$ appearing in Eq.~\eqref{eq:Lipschitz} is referred to as the Lipschitz constant of $f$.
	
	\newcommand{\indv}[1]{\mathbf{#1}}
	We will apply Lemma \ref{thm:Levy} to the scalar functional
	\begin{align}
		f_{\indv{i}\indv{j}}(U) \coloneqq \braket{\indv{i}|\mathcal{E}^{(k)}[U]|\indv{j}} = \sum_m \frac{\prod_{l=1}^k \braket{i^{(l)}|F_m|j^{(l)}} }{\Tr[F_m]^{k-1}}
	\end{align}
	where $\ket{\indv{i}} = \ket{i^{(1)}} \otimes \cdots \otimes \ket{i^{(k)}}$ is a state in the $k$-fold replicated space, and each $i^{(l)}$ runs over a basis for the Hilbert space of $S$. First we need to compute the Lipschitz constant of $f_{\indv{i} \indv{j}}$. This will proceed somewhat differently to the arguments of Ref.~\cite{Cotler2021}.
	
	To bound the left hand side of \eqref{eq:Lipschitz}, we define a parametrization of matrices $U(t) = (1-t)U_1 + t U_2$, which lies within the convex hull of unitary matrices for $t \in [0,1]$. We then have
	\begin{align}
		\frac{|f(U_1) - f(U_2)|}{\|U_1 - U_2\|_2} = \frac{\left| \int_0^1 \dif t \frac{\dif f(U(t))}{\dif t}\right|}{\left\|\int_0^1 \dif t \dot{U} \right\|_2} = \frac{\left| \int_0^1 \dif t \Tr\left[\frac{\partial f}{\partial U}\dot{U}\right] + \Tr\left[ \frac{\partial f}{\partial U^\dagger}\dot{U^\dagger}\right] \right|}{\left\|\int_0^1 \dif t \dot{U} \right\|_2} \leq 2 \max_U \left\|\frac{\partial f}{\partial U}\right\|_2 
	\end{align}
	where $\dot{U} = \dif U/\dif t$, and the maximum is taken over all $U$ in the convex hull of $\mathrm{U}(d)$. In the last step, we have used the Cauchy-Schwatz inequality applied to the Hilbert-Schmidt norm $|\Tr[A B]| \leq \|A\|_2 \|B\|_2$, along with the constancy of $\dot{U}$ and the relation $\| \partial f / \partial U\|_2 = \| \partial f / \partial U^\dagger\|_2$. Note that our definition of the matrix derivative is $(\partial f/\partial U)_{ab} = \partial f/U_{ba}$ and $(\partial f/\partial U^\dagger)_{ab} = \partial f/\bar{U}_{ab}$, where the variables $U_{ab}$ and $\bar{U}_{ab}$ are treated as being independent.
	
	Using the expression $F_m = \braket{0_A|U^\dagger|m}\braket{m|U|0_A}$ [Eq.~\eqref{eq:POVMOp}], the norm of the matrix derivative can be evaluted
	\begin{align}
		\left\| \frac{\partial f_{\indv{i}\indv{j}} }{\partial U} \right\|_2 = \left\| X + Y \right\|_2 \leq \|X\|_2 + \|Y\|_2
	\end{align}
	where we have defined
	\begin{align}
		X &= \sum_m \sum_{l = 1}^k \frac{\prod_{l'\neq l} \braket{i^{(l')}|F_m|j^{(l')}} }{\Tr[F_m]^{k-1}} \braket{i^{(l)}\otimes 0_A|U^\dagger|m} \ket{j^{(l)} \otimes 0_A}\bra{m} \\
		Y &= (k-1) \sum_m \frac{\prod_{l=1}^k \braket{i^{(l')}|F_m|j^{(l')}} }{\Tr[F_m]^k} (\ket{0_A}\bra{0_A} \otimes \mathbbm{I}_S) U^\dagger \ket{m}\bra{m}
	\end{align}
	Now recalling the definition $\tilde{F}_m = F_m/\Tr[F_m]$, we can express the squared norms $\|X\|_2^2 = \Tr[X^\dagger X]$ in terms of matrix elements of POVM operators
	\begin{align}
		\|X\|_2^2 &= \sum_m \sum_{l=1}^k\sum_{p=1}^k \delta_{j^{(l)} j^{(p)}} \left[\prod_{l'\neq l} \braket{i^{(l')}|\tilde{F}_m|j^{(l')} }\right] \left[\prod_{p'\neq p} \braket{j^{(p')}|\tilde{F}_m|i^{(p')} }\right] \Tr[F_m] \braket{i^{(l)}|\tilde{F}_m|i^{(p)}} \nonumber\\
		&= d_S\sum_{l=1}^k\sum_{p=1}^k \delta_{j^{(l)} j^{(p)}} \braket{\indv{i}_{\bar{l}} \otimes \indv{j}_{\bar{p}} \otimes i^{(l)}| \mathcal{E}^{(2k-1)}|\indv{j}_{\bar{l}} \otimes \indv{i}_{\bar{p}} \otimes i^{(p)}}
		\label{eq:NormX}
	\end{align}
	where $\ket{\indv{i}_{\bar{l}}}$ is a tensor product of all $\ket{i^{(l')}}$ for $l' \neq l$. Here we have invoked the representation of $\mathcal{E}^{(2k-1)}$ in terms of the POVM operators; see the right hand side of Eq.~\eqref{eq:Ensemble}. It is important to remember that $U$ here can be any matrix in the convex hull of $\mathrm{U}(d)$, i.e.~$U = \sum_i \lambda_i V_i$ with $\lambda_i \geq 0$,  $V_i^\dagger V_i = \mathbbm{I}$, and $\sum_i \lambda_i = 1$. This set is equal to the space of $d \times d$ complex matrices satisfying $\| U \|_\infty \leq 1$. We can still use the form $\mathcal{E}^{(2k-1)} = \sum_m q_m (\ket{\phi_m}\bra{\phi_m})^{\otimes k}$, where $\ket{\phi_m}$ are normalized wavefunctions, and $q_m = \braket{m|U (\mathbbm{I}_S \otimes \ket{0_A}\bra{0_A}) U^\dagger|m}/d_S$. We will make use of the following matrix inequality
	\begin{align}
		\mathcal{E}^{(1)} &\equiv \frac{1}{d_S} \sum_m \braket{0_A|U^\dagger|m}\braket{m|U|0_A} \leq \mathbbm{I}/d_S & \text{for any }U \text{ such that } \|U\|_\infty \leq 1
		\label{eq:IneqE1}
	\end{align}
	which follows straightforwardly from the fact that $\braket{\phi|\mathcal{E}^{(1)}|\phi} = \braket{\phi\otimes 0_A|U^\dagger U|\phi \otimes 0_A}/d_S \leq \braket{\phi|\phi} \|U^\dagger U\|_\infty/d_S \leq \braket{\phi|\phi}/d_S$.
	
	
	Now, since $\mathcal{E}^{(2k-1)}$ is a positive operator, we have $|\braket{a|\mathcal{E}^{(2k-1)}|b}| \leq \sqrt{\braket{a|\mathcal{E}^{(2k-1)}|a}\braket{b|\mathcal{E}^{(2k-1)}|b}}$ for any $\ket{a}, \ket{b}$ in the replicated Hilbert space. After applying this to the summand in \eqref{eq:NormX}, we then use 
	\begin{align}
		\braket{\indv{i}_{\bar{l}} \otimes \indv{j}_{\bar{p}} \otimes i^{(l)}|\mathcal{E}^{(2k-1)}|\indv{i}_{\bar{l}} \otimes \indv{j}_{\bar{p}} \otimes i^{(l)}} &= \sum_m q_m  |\braket{i^{(l)}|\phi_m}|^2\prod_{l'\neq l}|\braket{\phi_m|i^{(l')}}|^2 \prod_{p'\neq p}|\braket{\phi_m|i^{(p')}}|^2 \nonumber\\
		&\leq \sum_m q_m |\braket{i^{(l)}|\phi_m}|^2 = \braket{i^{(l)}|\mathcal{E}^{(1)}|i^{(l)}} \leq \frac{1}{d_S}
	\end{align}
	where in the last step we have used Eq.~\eqref{eq:IneqE1}. This implies that the summand of \eqref{eq:IneqE1} have modulus at most 1, and so summing over $l$, $p$ we finally obtain $\|X\|_2^2 \leq k^2$.
	
	Similarly, we can express $\|Y\|_2^2$ in terms of the moments $\mathcal{E}^{(2k)}$:
	\begin{align}
		\|Y\|_2^2 &= (k-1)^2 \sum_m \left|\prod_{l=1}^k \braket{i^{(l)}|\tilde{F}_m|j^{(l)}} \right|^2 \Tr[F_m] \nonumber\\
		&= (k-1)^2  \braket{\indv{i} \otimes \indv{j}|\mathcal{E}^{(2k)}|\indv{j} \otimes \indv{i}}
	\end{align}
	Following a similar line of reasoning as before, we have $ \braket{\indv{i} \otimes \indv{j}|\mathcal{E}^{(2k)}|\indv{j} \otimes \indv{i}} \leq 1$. Hence $\|Y \|_2^2 \leq (k-1)^2$. Putting everything together, we get
	\begin{align}
		\frac{|f_{\indv{i}\indv{j}}(U_1) - f_{\indv{i}\indv{j}}(U_2)|}{\|U_1 - U_2\|_2} \leq \eta \coloneqq 2k - 1
	\end{align}
	The rest of the proof follows the same logic as Ref.~\cite{Cotler2021}. Defining $\Delta^{(k)} \coloneqq \mathcal{E}^{(k)}[U] - \mathcal{E}^{(k)}_{\rm Haar}$, we have $\|\Delta^{(k)}\|_1 \leq \sum_{\indv{i}, \indv{j}} |\Delta^{(k)}_{\indv{i} \indv{j}}|$. Therefore the probability that the trace distance exceeds some value $\epsilon$ can be upper bounded
	\begin{align}
		\mathbbm{P}_{U \sim \textrm{Haar}}\left( \big\|\mathcal{E}^{(k)}[U] - \mathcal{E}^{(k)}_{\rm Haar}\big\|_1 \geq \epsilon \right) \leq  \mathbbm{P}_{U \sim \textrm{Haar}}\left( \sum_{\indv{i}, \indv{j}} |\Delta^{(k)}_{\indv{i} \indv{j}}| \geq \epsilon \right) \leq \mathbbm{P}_{U \sim \textrm{Haar}}\left( |\Delta^{(k)}_{\indv{i} \indv{j}}| \geq \epsilon/d^{2k}_S \textrm{ for some }\indv{i}, \indv{j}\right)
	\end{align}
	Employing Lemma \ref{thm:Levy} and taking a union bound, we have
	\begin{align}
		\mathbbm{P}_{U \sim \textrm{Haar}}\left( \big\|\mathcal{E}^{(k)}[U] - \mathcal{E}^{(k)}_{\rm Haar}\big\|_1 \geq \epsilon \right) \leq 4 d^{2k}_S \exp\left(-\frac{d\epsilon^2}{18\pi^3(2k-1)d_S^{4k}} \right)
	\end{align}
	For qubits we have $d = 2^{N_S + N_A}$ and $d_S = 2^{N_S}$, and so the tomographic ensemble forms an $\epsilon$-approximate $k$-design with probability $1-\delta$ whenever $\delta$ is less than the right hand side of the above, giving
	\begin{align}
		2^{N_A} \geq \frac{18 \pi^3(2k-1) 2^{4kN_S - 1} }{\epsilon^2}\Big( 2k N_S \log 2 + \log(4/\delta)\Big)
	\end{align}
	This is achieved whenever $N_A$ scales according to the relation quoted in Eq.~\eqref{eq:HaarScaling}.\hfill $\square$ 
	
	\subsection{Dual-unitary circuit}
	
	The second case where the existence of $k$-designs can be rigorously proven is when the unitary $U$ is a brickwork circuit made up of two-site gates that are each dual-unitary. In this section, we will allow for arbitrary local Hilbert space dimension $q \geq 2$, i.e.~we consider systems of $N$ qudits; accordingly, measurement outcomes $m$ can take $q$ distinct values.
	
	A unitary gate acting on two qudits can be viewed as a rank-4 tensor $U^{o_1 o_2}_{i_1 i_2}$, with two indices $i_{1,2}$ for the initial state of the qudits and two indices $o_{1,2}$ for the corresponding outputs. The gate is dual-unitary if the components of this tensor also describe a unitary matrix when viewed as a map from inputs $i_1 o_1$ to outputs $i_2 o_2$. This is illustrated in Fig.~\ref{fig:GraphNot}(a). The space of dual-unitary gates acting on qubits ($q = 2$) was classified in Ref.~\cite{Bertini2018}. Brickwork circuits made up of dual-unitary gates describe a form of many-body quantum dynamics wherein many properties can be calculated exactly, such as two-point correlation functions, entanglement entropies, and out-of-time-order correlators \cite{Bertini2019,Gopalakrishnan2019,Bertini2019a}. 
	
	\begin{figure}
		\centering
		\includegraphics[width=510pt]{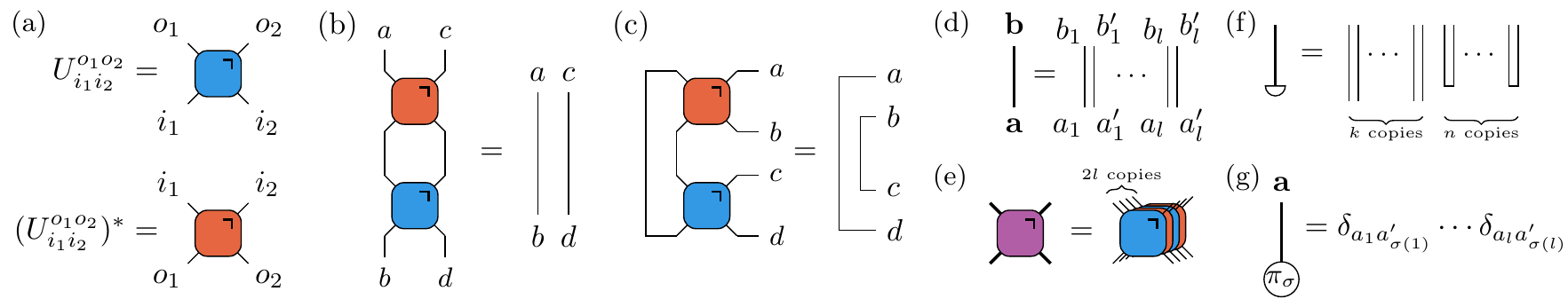}
		\caption{Summary of graphical notation used here, following the same conventions as Ref.~\cite{Claeys2022}. (a) A two-qudit unitary and its complex conjugate are represented as blue and red tensors, respectively. (b) Representation of the unitary condition. (c) Representation of the dual-unitary condition. (d,e) When $l$-fold replicas are constructed, as in Eq.~\eqref{eq:GeneralizedEnsemble}, thick lines are used to represent the $l$ forward indices $a_i$ and $l$ reverse indices $a_i'$. Time evolution in the replicated space can be expressed using multiple copies of each unitary and its complex conjugate. (f) In Eq.~\eqref{eq:GeneralizedEnsemble}, $n$ copies of the unnormalized POVM operator $F_m$ are traced over, while $k$ copies are left untouched; this operation is denoted using a semicircle. (g) The permutation tensor $\pi_\sigma$ pairs each forward index $a_i$ with the reverse index $a'_{\sigma(i)}$ associated via the permutation $\sigma \in \Sigma_l$, where $\Sigma_l$ is the group of permutations of $l$ elements.}
		\label{fig:GraphNot}
	\end{figure}
	
	Recently it has been shown that under certain conditions, a many-body wavefunction $\ket{\Psi^{SA}} = U\ket{0}$ generated by time evolution under a dual-unitary circuit $U$ realises an exact $k$-design in its projected ensemble, in the limit of an infinite number of ancilla qudits $N_A \rightarrow \infty$ \cite{Ho2022,Claeys2022,Ippoliti2022}. In order for this result to hold, the dual-unitary must not be fine-tuned to an integrable point, and the initial state $\ket{0}$ and measurement basis $\{\ket{m}\}$ must form a solvable measurement scheme (using the terminology of \cite{Claeys2022}), meaning that certain criteria that depend on the circuit in question must be met. We will leverage results that were proved in this context to show that the tomographic ensemble formed from such a unitary will also form exact $k$ designs in the limit $N_A \rightarrow \infty$.
	
	By analogy to the moments of the projected ensemble \cite{Claeys2022}, here the moments of the tomographic ensemble can be studied analytically using a replica trick. We generalize the definition of the ensemble moments \eqref{eq:Ensemble} to
	\begin{align}
		\mathcal{E}^{(n,k)} = \sum_m q_m^{n+k} \big(\ket{\phi_m}\bra{\phi_m}\big)^{\otimes k} = \sum_m \big(\!\Tr[F_m]\,\big)^n F_m^{\otimes k}
		\label{eq:GeneralizedEnsemble}
	\end{align}
	where $F_m$ are the unnormalized POVM operators defined in Eq.~\eqref{eq:POVMOp}. This representation can be generalized to any number of layers $t$ ($t = 4$ layers are shown above). The properly normalized $k$th moment is recovered by analytically continuing $n$ and taking the replica limit $n \rightarrow 1-k$.
	
	Each term in the sum in \eqref{eq:GeneralizedEnsemble} can be expressed in terms of an $l$-fold copy of the original circuit $\mathcal{U} \coloneqq (U \otimes U^*)^{\otimes l}$, where $l = n+k$. In all copies, the initial state of the ancillas are the same state $\ket{0_A}$, and the final states of all qudits are projected onto $\ket{m}\bra{m}$. For $n$ of the copies, the inputs to the unitary in the region $S$ are traced out, while for the remaining $k$ copies, those inputs are left as free, constituting the components of the summand in \eqref{eq:GeneralizedEnsemble}. We will focus on initial states and measurement bases that are product states here. The summand can be graphically represented using the notation described in Fig.~\ref{fig:GraphNot} as
	\begin{align}
		\big(\!\Tr[F_m]\,\big)^n F_m^{\otimes k} = \vcenter{\hbox{\includegraphics{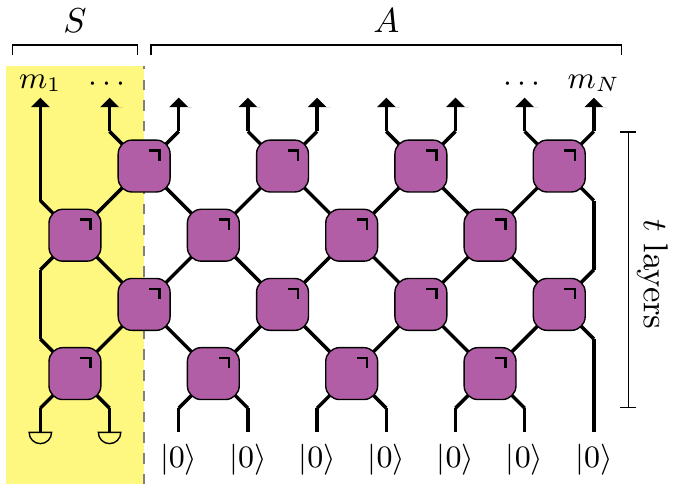}}}
		\label{eq:GeneralizedSummand}
	\end{align}
	We have highlighted the region corresponding to qudits in $S$ to distinguish this part of the circuit from the part acting on ancillas $A$. The latter part of the diagram will simplify upon taking the limit $N_A \rightarrow \infty$.
	
	Given that the initial state and measurement basis are product states, we will need to assume an additional property of $U$ which ensures that the measurement scheme is solvable. This property is found in the kicked Ising model \cite{Bertini2018}, as well as a family of gates introduced in Ref.~\cite{Claeys2022}; we refer interested readers to that work for details. Here, we will simply state this property, and assume it in the following. For any computational basis states $m_1$, $m_2$, the two-site gates we consider here must satisfy
	\begin{align}
		\vcenter{\hbox{\includegraphics[scale=1]{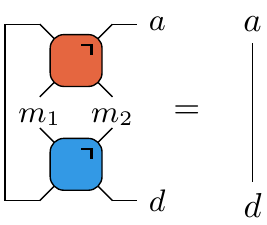}}}
		\label{eq:Solvable}
	\end{align}
	When this property is obeyed, the part of the circuit \eqref{eq:GeneralizedSummand} that acts on ancilla qudits simplified considerably in the limit $N_A \rightarrow \infty$ \cite{Claeys2022}
	\begin{align}
		q^{2(l-1)N_A} \times\sum_{m_1,\, \ldots\, m_{N_A}} \vcenter{\hbox{\includegraphics[scale=1]{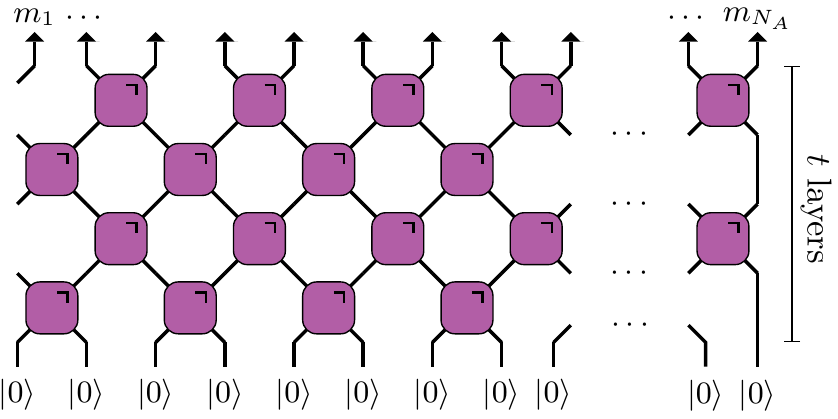}}}
		\hspace{5pt}\xrightarrow[N_A \rightarrow \infty]{} \hspace{5pt} \text{const.}\times\sum_{m_1}\sum_{\sigma \in \Sigma_l} \vcenter{\hbox{\includegraphics[scale=1]{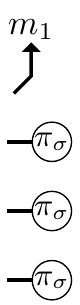}}}
		\label{eq:AncillaPermutations}
	\end{align}
	where the components of the rank-1 tensors $\pi_\sigma$, defined for each element of the permutation group of $l$ objects $\sigma \in \Sigma_l$, are given in Fig.~\ref{fig:GraphNot}(g). The above holds for any integer $k$ and any even number of layers $t \geq 2$, provided that the circuit is not integrable. An analogous result for odd $t$ can also be obtained, with different boundary conditions at the top. The factor of $q^{2(l-1)N_A}$ is required to ensure that the left hand side is finite and bounded in the limit $N_A \rightarrow \infty$.
	
	Generalized moments \eqref{eq:GeneralizedEnsemble} of the projected ensemble of the wavefunction $\ket{\Psi^{SA}} = U\ket{0}$ can be computed with the help of Eq.~\eqref{eq:AncillaPermutations}. By analytically continuing to $n \rightarrow 1-k$, the properly normalized moments can be obtained. When the number of layers $t$ is at least as large as $N_S$, one finds that the projected ensemble forms an exact $k$-design \cite{Ho2022,Claeys2022,Ippoliti2022}. We will use similar arguments to show that the tomographic ensemble also forms a $k$-design for $t \geq N_S$. A key ingredient will be the following relations
	\begin{align}
		\vcenter{\hbox{\includegraphics[scale=1]{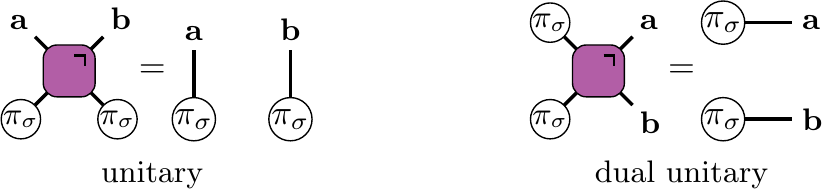}}}
		\label{eq:RepUnitary}
	\end{align}
	for any permutation $\sigma$. These follow from the unitarity and dual-unitarity of the two-qudit gates. In addition, the solvable measurement scheme condition \eqref{eq:Solvable} can be rewritten in the replicated space as
	\begin{align}
		\vcenter{\hbox{\includegraphics[scale=1]{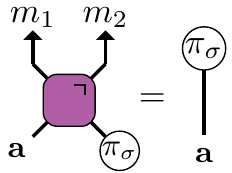}}}
		\label{eq:RepSolvable}
	\end{align}
	which will be made use of in the following.
	
	The generalized moment is obtained by summing Eq.~\eqref{eq:GeneralizedSummand} over $m$ and applying \eqref{eq:AncillaPermutations}, which leads to a significant simplification, and subsequent application of Eqs.~(\ref{eq:RepUnitary}, \ref{eq:RepSolvable}) allows further reduction (a representative case $t = 4$, $N_S = 2$ is shown in the following diagrams, but the arguments steps are readily generalised)
	\begin{align}
		\mathcal{E}^{(n,k)} &\propto \sum_{m_1 m_2 m_3} \sum_{\sigma \in \Sigma_l} \vcenter{\hbox{\includegraphics[scale=1]{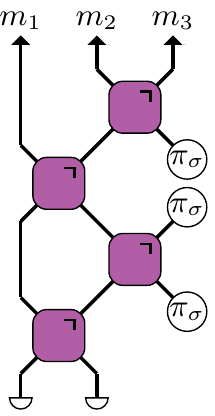}}} = \sum_{m_1 m_2 m_3} \sum_{\sigma \in \Sigma_l} \vcenter{\hbox{\includegraphics[scale=1]{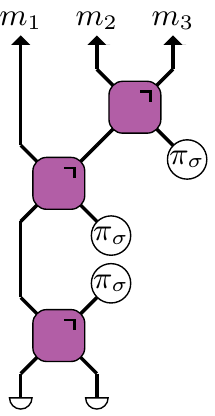}}} = \sum_{m_1} \sum_{\sigma \in \Sigma_l} \vcenter{\hbox{\includegraphics[scale=1]{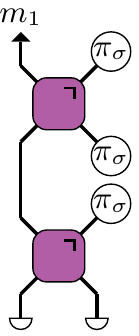}}}
		= \sum_{\sigma \in \Sigma_l} \vcenter{\hbox{\includegraphics[scale=1]{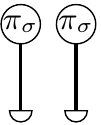}}}
	\end{align}
	Finally, by performing the trace over the $n$ copies [see Fig.~\ref{fig:GraphNot}(f)], we find that the generalized moment is an equal-weight sum of all permutation tensors over $k$ elements $\mathcal{E}^{(n,k)} \propto \sum_{\sigma \in \Sigma_k} \pi_\sigma$. The dependence on $n$ is then entirely through a constant of proportionality, and so the analytic continuation $n \rightarrow 1-k$ is readily taken. In fact, the value of this proportionality constant in the replica limit is fixed by the condition that $\Tr[\mathcal{E}^{(k)}] = 1$. Putting everything together, and noting that the sum over all permutation operators in the $k$-fold replica space is precisely the $k$th moment of the Haar ensemble, we conclude that the tomographic ensemble forms an exact $k$-design for all $k$. \hfill $\square$
	
	\section{Including randomness}
	
	In the main text, we described a `semi-randomized' version of our original protocol, where in each repetition of the experiment, $U$ is chosen at random from some distribution of unitaries that are all native to the quantum device in question. Here we describe how this modification reduces the systematic error of the estimators, and reduces the overheads in terms of number of ancillas.
	
	In this section, we will make reference to a scenario where the unitary is sampled from a discrete probability distribution $\{(p_c, U_c)\}$, where $c = 1, \ldots, C$ labels distinct unitaries $U_c$, which occur with probability $p_c$  (note however that all results can be straightforwardly generalized to continuous probability distributions). When the unitary $U$ itself is decided by a random process, we can write down a POVM $\{F_{m,c}\}$ such that the probability of both choosing a specific unitary $U_c$, and obtaining the outcome $m$ is $\Tr[\rho_S F_{m,c}]$. Specifically, we have
	\begin{align}
		F_{m,c} = p_c \times F_{m|U_c}
	\end{align}
	where $F_{m|U_c}$ is the POVM operator \eqref{eq:POVMOp} with $U$ replaced by $U_c$. Since each $\{F_{m|U_c}\}_m$ is itself a POVM, we have $\sum_{m,c} F_{m,c} = \mathbbm{I}$ as desired. Note that this formalism could be used to capture classical shadow tomography, where $U_c$ is sampled from the appropriate distribution of unitaries (global Clifford circuits or local Pauli rotations). The difference here is that a state design is approximately formed for each $U_c$, whereas in shadow tomography the POVM for a single unitary is far from being a $k$-design for $k \geq 2$, since the number of different possible measurement outcomes is not big enough ($2^{N_S}$ compared to the minimum number $4^N$ required to form a 2-design). 
	
	Now, we observe that the $k$th moments of the full POVM $F_{m,c}$ [Eq.~\eqref{eq:Ensemble}] are convex combinations of the moments for each individual POVM $F_{m|U_c}$, namely
	\begin{align}
		\mathcal{E}^{(k)}_{\rm full} \coloneqq \sum_{c=1}^C \sum_m \Tr[F_{m,c}] (\tilde{F}_{m,c})^{\otimes k} = \sum_{c=1}^C p_c \mathcal{E}^{(k)}[F_{m|U_c}]
		\label{eq:MomentsSemiRand}
	\end{align}
	where $\mathcal{E}^{(k)}[F_{m|U_c}]$ is the $k$th moments of the tomographic ensemble for the POVM \eqref{eq:POVMOp} where the unitary $U_c$ is used. If each separate POVM $\{F_{m|U_c}\}_m$ deviates from a perfect 2-design by an amount $\Delta_c$ (as quantified by the trace distance $\frac{1}{2}\|\mathcal{E}^{(2)} - \mathcal{E}_{\rm Haar}^{(2)}\|_1$), then by the triangle inequality $\mathcal{E}_{\rm full}^{(2)}$ will deviate from a state design by \textit{at most} the average value of the trace distance $\sum_c p_c \Delta_c$. Roughly speaking, if the projectors that make up each $F_{m|U_c}$ are uncorrelated with one another, then we expect that the differences $\mathcal{E}^{(2)}[F_{m|U_c}] - \mathcal{E}_{\rm Haar}^{(2)}$ will typically be in different directions in operator space, and so the errors will compound sub-additively, yielding a smaller value of the trace distance. Since the discrepancy between the $k = 2$ moments of the tomographic ensemble and the corresponding Haar moments governs the systematic error of the estimators, we see that adding randomness allows one to reduce any such error in our protocol.
	
	\begin{figure}
		\centering
		\includegraphics[width=0.5\linewidth]{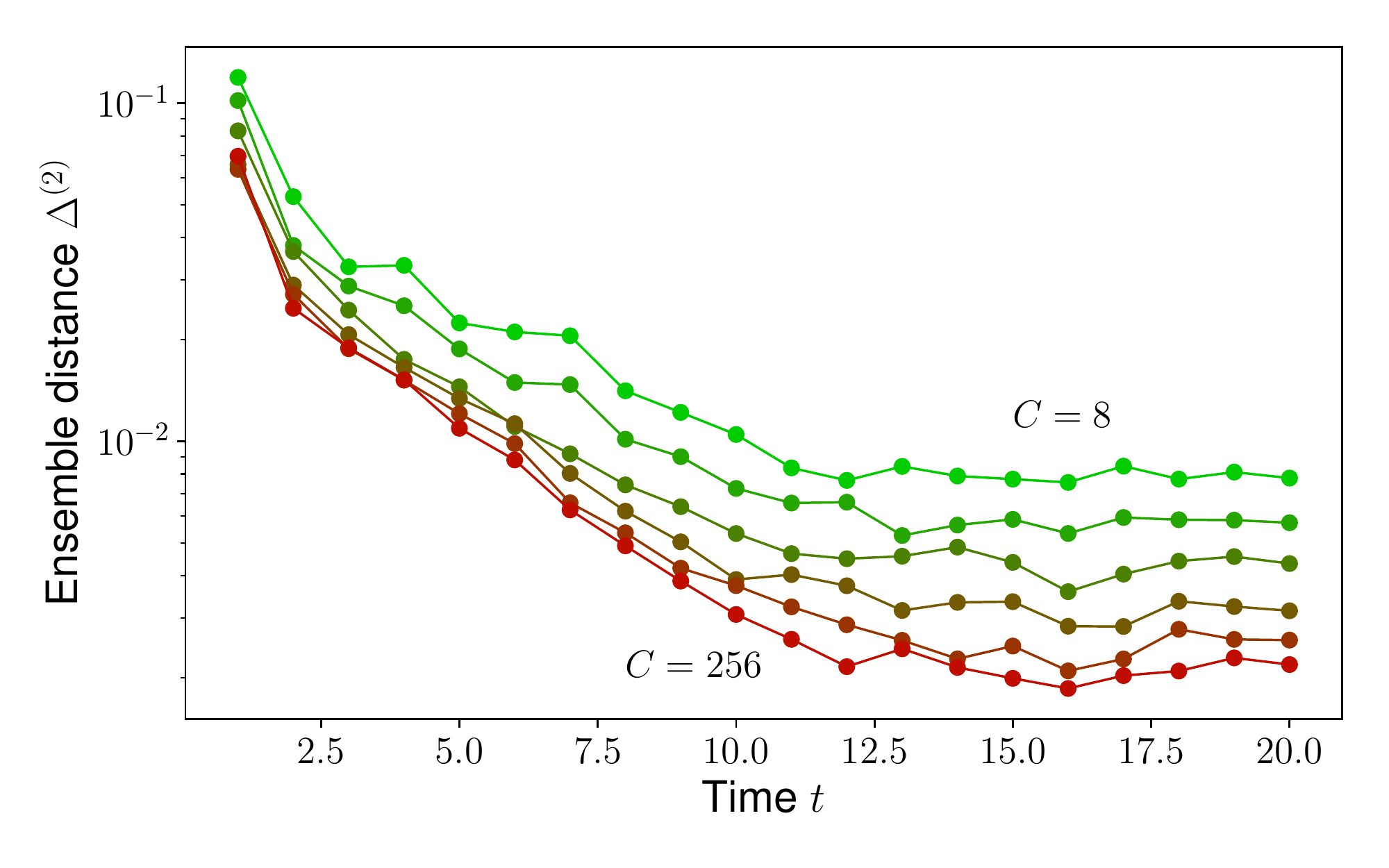}
		\caption{Trace distance $\Delta^{(2)}$ of the tomographic ensemble where the unitary is chosen at random from a discrete probability distribution $\{(p_c, U_c)\}$, where $c = 1, \ldots, C$, and the moments formed as in Eq.~\eqref{eq:MomentsSemiRand}. We vary the number of different unitaries $C$, from the smallest value $C = 8$ (green), doubling each time to $C = 16, 32, 64, \ldots$, up to $C = 256$ (purple). We always choose a uniform distribution for $p_c = 1/C$, and each unitary $U_c$ is generated by the tilted-field Ising Hamiltonian, as discussed in the main text. Increasing $C$ leads to better convergence towards a $k$-design---in particular, after a $C$-independent time, $\Delta^{(2)}$ saturates at a value proportional to $C^{-1/2}$.} 
		\label{fig:my_label}
	\end{figure}
	
	To verify that the moments of the full ensemble $\mathcal{E}^{(k)}_{\rm full}$ are indeed closer to being a $k$-design than each separate POVM, we compute the $k = 2$ trace distance for a particular family of different distributions of unitaries $\{(p_c, U_c)\}$. In each case, we keep the probabilities uniform $p_c = 1/C$, and vary the number of different unitaries $C$. Each $U_c$ is generated by the tilted-field Ising model Hamiltonian discussed in the main text, $H(t) = \sum_j X_j X_{j+1} + h^x(t)X_j +h^y(t) Y_j$. The values of $h^{x,y}(t)$ change abruptly every $\tau = 0.5$ time units, and each unitary $U_c$ has a different sequence of field values. Before running any simulations, we choose the actual field values for each $c$ through independent random sampling from a normal distribution with zero mean and standard deviation $\sqrt{2}$, and use these field values to construct the moments \eqref{eq:MomentsSemiRand}. In the limit $C \rightarrow \infty$, this describes the protocol used to generate the data shown in Fig.~\ref{fig:benchmark}. We see that increasing $C$ does indeed reduce the trace distance, and empirically we find that the plateau value of $\Delta^{(2)}$ scales as $1/\sqrt{C}$. Notably, this is the behaviour that we would expect if we assumed that every wavefunction $\ket{\phi_{m,c}}$ in the ensemble was an independent randomly distributed vector. Of course, in the true $C \rightarrow \infty$ limit, one can find two possible unitaries $U_c$, $U_{c'}$ that are very close to one another such that this assumption of independence will fail; thus at large enough $C$ the trace distance should saturate to a finite (but very small) value. Since constructing the tomographic ensemble for large values of $C$ is computationally demanding, we find it easier to properly assess the performance of this randomized protocol by simulating the whole procedure, as in the data presented in Fig.~\ref{fig:benchmark}. \\
	
	Note that adding randomness does not increase the classical computational overhead because the inverse map $\mathcal{M}^{-1}$ is chosen to be the universal form \eqref{eq:MInverse}. If, on the other hand, we were to compute $\mathcal{M}^{-1}$ exactly, we would have to compute $2^N \times C$ individual wavefunctions, where $C$ is the number of different unitaries in the distribution. Also, from our simulations we find that the full ensemble can approach a 2-design very closely even with a modest number of ancillas, because trace distance for each separate POVM $\{F_{m|U_c}\}_m$ does not need to be particular small, provided that we can sample from a sufficiently diverse range of unitaries.
	
	
	\section{Accounting for symmetries}
	
	Some quantum simulators possess intrinsic symmetries that cannot be readily broken, e.g.~number conservation in ultracold atomic gases. This restricts the space of unitaries $U$ that are available, which in turn leads to constraints on the POVMs that can be realised with our protocol. Focusing on Abelian symmetry groups, in this section we will show that the moments of the tomographic ensemble tend towards different universal form that respects this symmetry: Specifically, within each symmetry charge sector an approximate state design is formed. Again this occurs provided that $U$ is sufficiently entangling, and, in the case where the symmetry is continuous, the initial state of the ancillas must also have a small enough effective chemical potential (i.e.~$\ket{0}_A$ is not close to being a maximum- or minimum-charge state).
	
	The Hilbert space of a system that respects some Abelian symmetry can be decomposed into charge sectors $q$, spanned by orthogonal projectors $P_q$, which are not coupled by symmetric unitaries: $P_q U P_{q'} = 0$ for $q \neq q'$. Any target density matrix $\rho_S$ that can be prepared using symmetric operators will also be constrained to have vanishing coherences between different charge sectors, i.e.~$[\rho_S, P_q] = 0$. (In the nomenclature of Ref.~\cite{Buca2012}, this corresponds to a `weak symmetry', in contrast to a state that has support in only one charge sector, which is `strongly symmetric'.) Therefore, only charge-diagonal observables and states need be considered, since operators that couple different charge sectors have vanishing expectation values. Symbolically, we have
	\begin{align}
		\oper &= \bigoplus_q \oper_q & \rho_S &= \bigoplus_q \rho_q.
		\label{eq:OperChargeDecomp}
	\end{align}
	
	We naturally presume that computational basis states and the ancilla initial states each have definite charge. Therefore each POVM operator $F_m$ [Eq.~\eqref{eq:POVMOp}] will lie in a particular charge sector $q(m)$, where $q(m)$ is the charge of the system that is required to match the total system plus ancilla charge before applying $U$ to the final measured charge $m$. This constraint prevents the formation of full state designs, since superpositions of states with different charges are forbidden. Instead, we find that generic symmetry-respecting dynamics yields a POVM for which the tomographic ensemble forms an approximate state design within each charge sector---we call such a distribution a `block-diagonal state design'. That is, we can decompose $\mathcal{E}^{(k)} = \bigoplus_q \mathcal{E}^{(k)}_q$, where $\mathcal{E}^{(k)}_q$ contains only terms in Eq.~\eqref{eq:Ensemble} for which $q(m) = q$, and we find that $\mathcal{E}^{(k)}_q$ approaches the $k$th moment of the Haar ensemble over $\text{span}(P_q)$. Note that the block-diagonal Haar ensemble is the distribution that maximizes randomness subject to the constraints imposed by symmetry.
	
	Provided that $\mathcal{E}^{(2)}$ takes this universal form, expectation values of charge-diagonal operators can be estimated using \eqref{eq:InvConstruct}, after replacing the expression in Eq.~\eqref{eq:MInverse} with a block-diagonal superoperator
	\begin{align}
		\mathcal{M}^{-1} &= \bigoplus_q \mathcal{M}^{-1}_q & \text{where }\mathcal{M}^{-1}_q[\oper_q] = (d_q + 1)\oper_q - \Tr[\oper_q]P_q,
		\label{eq:MInverseSymmetry}
	\end{align}
	where $\oper_q$ is the submatrix of the operator $\oper$ contained within the charge sector $q$, as in Eq.~\eqref{eq:OperChargeDecomp}.\\
	
	It is relatively straightforward to see that if the moments of the tomographic ensemble do converge to a universal form, then it must be the block-diagonal state design described above. This follows from considering the behaviour of a unitary $U$ sampled from the maximally random distribution of charge-conserving unitaries, where each submatrix $U_q$ describing the behaviour of $U$ within charge sector $q$ is drawn from the Haar ensemble. By considering one block $q$ at a time, one can use the same method as in Eq.~\eqref{eq:HaarAverageMoment} to see that the mean value of $\mathcal{E}_q^{(k)}$ is equal to the $k$th moments of the Haar ensemble over the space spanned by $P_q$. The concentration of measure results given in the previous section can also be used to bound the deviation of a given symmetry sector from being a state design, with the Hilbert space dimensions $d$, $d_S$ replaced with their appropriate charge-restricted values: Specifically, $d_S$ should be replaced with the number of system states with a fixed charge $q$, and $d$ should be replaced with the number of measurement outcomes that could arise starting from a state where the system qubits have charge $q$, and the ancilla qubits have the charge determined by $\ket{0}_A$.
	
	For discrete symmetry groups, $d$ increases exponentially with system size for each charge block as before, and so state designs are formed with overwhelmingly high probability. However, if the symmetry is continuous (such that there is a conserved charge \textit{density}), then the effective dimension $d$ will scale much more slowly with system size when the charge of the ancilla initial state has near-maximum or near-minimum charge, reflecting the fact that there are fewer possible final measurement outcomes $m$, that are compatible with the initial charge configuration. Because of this, the ancilla initial state should be initialized with a non-extremal charge distribution if one is to expect formation of block-diagonal quantum state designs.\\
	
	\begin{figure}
		\centering
		\includegraphics{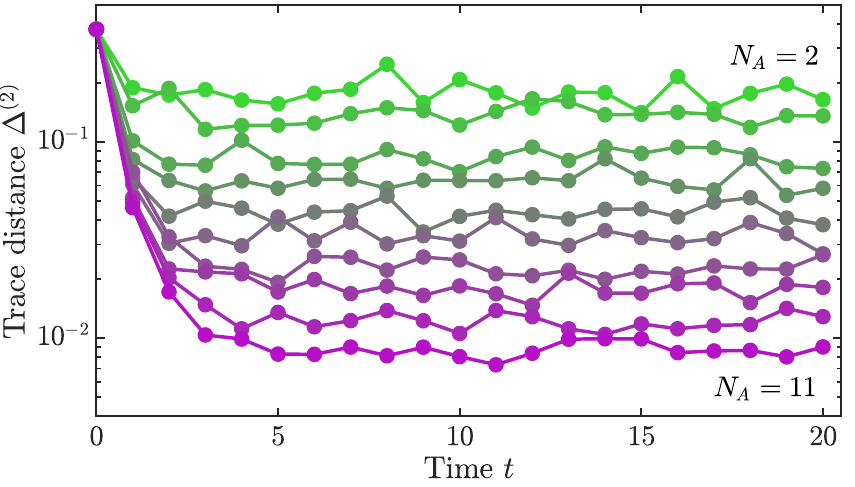}
		\caption{Trace distance between the $k = 2$ moments of the tomographic ensemble for dynamics generated by the XXZ model, and the fixed point distribution of states where wavefunctions are Haar-random within each symmetry charge block. We vary the parameters of the model periodically in time, such that the effective Floquet unitary describing evolution over one time period is $U_F = e^{-\iu \tau H_2} e^{-\iu \tau H_1}$, with $H_{1,2}$ both of the form \eqref{eq:XXZ}, and $\tau = 0.5$. The staggered field $h^z = 0.6$ throughout, while $\Delta = 0.8$ in $H_1$ and $\Delta = -1.7$ in $H_2$. Here, there are $N_2 = 3$ system qubits, located at the centre of a chain with open boundary conditions.}
		\label{fig:trace_dist_XXZ}
	\end{figure}
	
	We also provide numerical evidence that for representative symmetry-conserving unitaries, the tomographic ensemble approaches a block-diagonal state design. As a representative example, we consider dynamics generated by the XXZ Hamiltonian in a staggered longitudinal field
	\begin{align}
		H = \sum_j X_jX_{j+1} + Y_j Y_{j+1} + \Delta Z_j Z_{j+1} + (-1)^j h^z Z_j
		\label{eq:XXZ}
	\end{align}
	This model possesses a $\mathrm{U}(1)$ symmetry generated by operators $e^{\iu \theta \sum_j Z_j/2}$, which implements a rotation of all spins by an angle $\theta$ about the $z$-axis. The staggered field is included to break the integrability of this model, which allows for chaotic dynamics. Again we use Floquet evolution $U = U_F^t$, with $U_F = e^{-\iu H_2 \tau} e^{-\iu H_1 \tau}$, where $H_{1,2}$ have different values of the anisotropy parameter $\Delta$. In our simulations, we pick $\tau = 0.5$, and $h^z = 0.6$, $\Delta = 0.8$ in $H_1$, and $\Delta = -1.7$ in $H_2$. We have verified that qualitatively similar behaviour is seen for other choices of parameters.
	
	The initial state of the ancillas is chosen to be a staggered state $\bigotimes_j[\ket{0}_{2j-1} \otimes \ket{1}_{2j}]$. This state is chosen because there are a large number of states with the same total charge as this, compared to states that have near-extremal magnetization, i.e.~those that are close to all $\ket{0}$, or all $\ket{1}$. This ensures that there will be a large number of different possible measurement outcomes, which is necessary for the formation of a state design. We have found that a much larger number of ancillas are needed to form an approximate state design when the initial state has maximum magnetization.
	
	We compute the trace distance $\Delta^{(2)}$ between the $k = 2$ moments of the tomographic ensemble and the corresponding moments of the block-diagonal Haar ensemble. The results are plotted in Fig.~\ref{fig:trace_dist_XXZ}. Again we see similar trends to the trace distance for the tilted-field Ising model: After an initial transient period, the trace distance plateaus at a value that scales exponentially with the number of ancillas. This behaviour can be understood in the same way as before, by noting that the trace distance is a sum of contributions from each charge sector $q$, and that the number of measurement outcomes $m$ that reside in each sector is exponentially small in $N_A$ for all $q$ (assuming $N_A/N_S$ is large).\\
	
	In addition to conservation laws that are associated with unitary symmetries, one could in principle also consider conservation of energy due to time-translation symmetry. This applies when $U$ is generated from evolution under a time-independent Hamiltonian $U = e^{-\iu H t}$. However, from the results of Ref.~\cite{Cotler2021}, where the projective ensemble is studied, we anticipate that the evolution time required to reach the appropriate universal form will be much longer in this case. Since it is almost always possible to introduce some form of time-dependence in the Hamiltonian in experiments, we will not address this case here, instead leaving it to future work.
	
	
	\section{Details of classical post-processing}
	
	In this section we provide additional details on how properties of the target state $\rho_S$ can be estimated from experimentally observed measurement outcomes.

	\subsection{Optimality of the inverse map \eqref{eq:InvConstruct}}
	
	In the main text, we stated that the choice of inverse map $\mathcal{G}$ that minimizes the average-case variance is given by Eq.~\eqref{eq:InvConstruct}. Here we prove this statement. Our logic follows a similar line of reasoning to the arguments given in Ref.~\cite{Scott2006}, with the difference that here---in the spirit of shadow tomography \cite{Aaronson2018,Aaronson2019,Huang2020}---the goal is to estimate specific expectation values, rather than perform full tomography of the density matrix.\\

	For a given informationally complete POVM channel $\mathcal{F}$ [Eq.~\eqref{eq:POVMChannel}], a linear estimator $\hat{\theta}_\oper$ for any expectation value $\braket{\oper}$ can be represented as a dual vector $(w| \in \mathbbm{R}^{2^N*}$ satisfying $(w| \mathcal{F} = \llangle \oper|$. In particular, when we have a left inverse of $\mathcal{F}$, i.e.~an operator $\mathcal{G}$ satisfying $\mathcal{G}\mathcal{F} = \text{id}$, then we can set $(w| = \llangle \mathcal{O}|\mathcal{G}$. Thanks to this condition, experimental data can be processed to form a quantity $\hat{\theta}_\oper = M^{-1} \sum_{r = 1}^M (w|m^{(r)})$ (where $m^{(r)}$ is the measurement bitstring for repetition $r$) which is an unbiased estimator, since $\mathbbm{E} (w|m^{(r)}) = (w|\mathcal{F}|\rho_S\rrangle = \llangle \oper | \rho_S \rrangle \equiv \braket{\oper}$. While performance could in principle be improved by harnessing more sophisticated nonlinear estimation schemes, e.g.~maximum likelihood estimation, here we mainly analyse linear estimators. (One relatively simple example employed in Ref.~\cite{Huang2020} is to calculate a median-of-means, rather than just the mean used here, which reduces the chances of finding outliers.)
	
	Since $(w|$ is non-unique whenever $N_A > N_S$, we wish to find a choice that is optimal. However, because the variance depends on the state $\rho_S$ itself, which in principle is not known in advance. As in Ref.~\cite{Scott2006}, to reflect our lack of \textit{a priori} knowledge of $\rho_S$ we first average over all unitarily equivalent states, and then minimize this averaged variance. This amounts to minimization of
	\begin{align}
		\Delta[w] = \frac{1}{d_S} \sum_m w_m^2 \Tr[F_m]
	\end{align}
	subject to the constraint $(w|\mathcal{F} = \llangle \oper|$. We will show that the choice $(w^*| = \llangle \oper |\mathcal{G}^*$, where the inverse map $\mathcal{G}^*$ is given in Eq.~\eqref{eq:InvConstruct}, achieves this minimum.
	
	Firstly, we show that $(w^*|$ is a valid estimator. First observe that the channel $\mathcal{M}$ defined in Eq.~\eqref{eq:InvConstruct} can be written as either $\tilde{\mathcal{F}}^\dagger\mathcal{F}$ or $\mathcal{F}^\dagger\tilde{\mathcal{F}}$. Then, we have
	\begin{align}
		(w^*|\mathcal{F} = \llangle \oper|\mathcal{M}^{-1} \tilde{\mathcal{F}}^\dagger \mathcal{F} = \llangle \oper|\mathcal{M}^{-1} \mathcal{M} = \llangle \oper|
	\end{align}
	Using the above, we can establish that indeed $\mathbbm{E}_{\hat{m}}\hat{\theta}_\oper = \sum_m p_m(w^*|m) = (w^*|p) = (w^*|\mathcal{F}|\rho_S\rrangle = \llangle O|\rho_S\rrangle = \braket{\oper}$. By virtue of the above, any valid estimator $(w|$ can be written as $(w| = (w^*| + (b|$, where $(b|$ satisfies $(b|\mathcal{F} = 0$. Now we evaluate the functional
	\begin{align}
		\Delta[w] = \Delta[w^*] + \Delta[b] + \frac{2}{d_S} \sum_m \Tr[F_m] (w^*|m)(m|b) 
		\label{eq:DeltaW}
	\end{align}
	We now use the explicit form of $(w^*|$ given in \eqref{eq:InvConstruct} to obtain $(w^*|m) = \llangle \oper | \mathcal{M}^{-1}| F_m \rrangle/\Tr[F_m]$. The factors of $\Tr[F_m]$ then cancel in the summand in \eqref{eq:DeltaW}, allowing us to write
	\begin{align}
		\sum_m (w^*|m)(b|m) \Tr[F_m] &= \sum_m \llangle \oper | \mathcal{M}^{-1}|F_m\rrangle (m|b) \nonumber\\
		&= \sum_m \llangle \oper | \mathcal{M}^{-1} \mathcal{F}^\dagger |b)
	\end{align}
	Now, since $(b|\mathcal{F} = 0$ by definition, the sum in \eqref{eq:DeltaW} vanishes. Since $\Delta[b] \geq 0$ for any $(b| \in \mathbbm{R}^{2^N*}$ with equality if and only if $b = 0$, we conclude that $(w^*|$ achieves the global minimum of the functional $\Delta$, subject to the constraint of being a valid estimator. The value of this global minimum is
	\begin{align}
		\Delta[w^*] =  \frac{1}{d_S}\llangle \oper| \mathcal{M}^{-1} \left(\sum_m \Tr[F_m] |\tilde{F}_m\rrangle\llangle \tilde{F}_m|\right) \mathcal{M}^{-1} |\oper\rrangle = \frac{1}{d_S}\llangle \oper| \mathcal{M}^{-1} | \oper \rrangle.
		\label{eq:GlobalMin}
	\end{align}
	with a corresponding average variance of $\mathbbm{E}_{\rho_s} \Var \hat{\theta}_\oper|_{w^*} = M^{-1}(\Delta[w^*] - \mathbbm{E}_{\rho_S}\braket{\oper}^2)$. \hfill $\square$
	
	\subsection{Bounding the variance of estimators}
	
	As we showed in the previous section, the dual vector $(w_0|$ produces the estimator with the smallest possible variance averaged over all possible input states. However, this does not give us complete information about the variance that one would find for specific choices of $\rho_S$. Indeed, in principle there could exist particular adversarial input states for which the variance is exceptionally high, even if the average variance is small. To ascertain $\Var \hat{\theta}_\oper$ for any particular $\rho_S$, we can use Eq.~\eqref{eq:Variance3Design}, which is expressed in terms of the third moments of the tomographic ensemble $\mathcal{E}^{(3)}$ [see Eq.~\eqref{eq:Ensemble}]. Here, we will use this expression to obtain upper bounds for the variance of estimators of expectation values, as well as nonlinear properties of $\rho_S$.
	
	We are particularly interested in cases where the moments of the tomographic ensemble approach their universal maximum-randomness forms (the form in question depends on whether blocking is used or not, and whether any symmetries are present). In these cases, analytic expressions for $\mathcal{M}^{-1}$ and $\mathcal{E}^{(3)}$ can be obtained which allow the variance, treated as a joint functional of $\rho_S$ and $\oper$, to be specified explicitly. We aim to obtain simple upper bounds for the variance in such cases, focussing on setups with no conservation laws, either with or without blocking. As we will show, in the case with (without) blocking the functional form of the variance becomes the same as that of classical shadow tomography with random global (local) gates \cite{Huang2020}.
	
	To prove this, it will be helpful to represent conventional shadow tomography using the POVM formalism employed in this work. There, a unitary $U$ is applied to the system only (no ancillas are used), before measurement in the computational basis, giving an outcome $m_S$. The unitaries are randomly sampled from an appropriate discrete set $\mathcal{U}$ with probabilities $p_U$. We can then define POVM operators $F_{m_S, U}$ as
	\begin{align}
		F_{m_S, U} = p_U U^\dagger \ket{m_S}\bra{m_S}U
	\end{align}
	Each operator of the above form corresponds to an event where the unitary chosen is $U$, and the subsequent measurement outcome is $m$. Again, these are rank-1 projectors, and so moments of the tomographic ensemble can be formed:
	\begin{align}
		\mathcal{E}^{(k)} = \sum_{U \in \mathcal{U}} \sum_{m_S \in \{0,1\}^{N_S}} p_U \left(U^\dagger \ket{m_S}\bra{m_S} U\right)^{\otimes k}
		\label{eq:MomentRandom}
	\end{align}
	
	In addition to Eq.~\eqref{eq:Variance3Design}, we will also prove a useful result that allows us to express the variance of nonlinear estimators in terms of the third moment of the tomographic ensemble.
	
	\subsubsection{No blocking}
	
	In the protocol with a single collection of ancillas and no conservation laws, the tomographic ensemble approaches a $k$-design over all wavefunctions in the Hilbert space of $S$. The second and third moments therefore take their universal form
	\begin{align}
		\mathcal{E}^{(2)} &= \frac{1}{d_S(d_S+1)}(\mathbbm{I} + \pi_{(1 2)} ) \label{eq:Haar2}\\
		\mathcal{E}^{(3)} &= \frac{1}{d_S(d_S+1)(d_S+2)}\sum_{\sigma \in \Sigma_3} \pi_\sigma
		\label{eq:HaarLowMoments}
	\end{align}
	Because the POVM operators are rank-1 projectors $F_m = d_S q_m \ket{\phi_m}\bra{\phi_m}$, the second moment $\mathcal{E}^{(2)}$ fully specifies the map $\mathcal{M}$. Specifically, the two share the same matrix elements
	\begin{align}
		\braket{ij|\mathcal{E}^{(2)}|kl} = \frac{1}{d_S} \times \llangle E_{ik}| \mathcal{M}|E_{jl}\rrangle
		\label{eq:MatrixElements}
	\end{align}
	(The factor of $d_S^{-1}$ reflects the fact that the POVM operators sum to $\sum_m F_m = \mathbbm{I}$, which is a factor of $d_S$ larger than the first moment of a quantum state design $\mathcal{E}^{(1)} = \mathbbm{I}/d_S$.) Combining Eqs.~\eqref{eq:HaarLowMoments} and \eqref{eq:MatrixElements} gives $\mathcal{M} = (d_S+1)^{-1}[\text{id} + |\mathbbm{I}\rrangle \llangle \mathbbm{I}|]$, which is readily inverted, proving Eq.~\eqref{eq:MInverse}.
	
	We will compare our scheme to classical shadow tomography with random global Clifford gates. There, the set of unitaries $\mathcal{U}$ are all Clifford operations on $N_S$ qubits, and the probabilities are uniform $p_U = |\mathcal{U}|^{-1}$. Since the Clifford group forms a 3-design \cite{Webb2016}, the $k$th moments of the shadow tomographic POVMs [Eq.~\eqref{eq:MomentRandom}] are the same as those of the tomographic ensemble of our protocol for $k \leq 3$. The variance of the estimators $\hat{\theta}_\oper$ is fully determined by the third moments $\mathcal{E}^{(k \leq 3)}$, and so we can conclude that our scheme allows one to estimate expectation values to the same degree of uncertainty.
	
	Useful upper bounds for the variance in conventional shadow tomography are given in the supplement of Ref.~\cite{Huang2020}. In particular, we have
	\begin{align}
		\left.\Var \hat{\theta}_\oper \right|_{\text{3-design}} = \frac{1}{M} \frac{d_S+1}{d_S+2} \bigg(\!\Tr[\oper^2_0] + 2\Tr[\rho_S \oper^2_0]\,\bigg) \leq \frac{1}{M} \times 3 \| \oper_0 \|_2^2,
		\label{eq:VarBound3Design}
	\end{align}
	where $\oper_0 = \oper - \Tr[\oper]\mathbb{I}/d_S$ is the traceless part of $\oper$, and $\| C \|_2 = \sqrt{\Tr[C^\dagger C]}$ is the Frobenius norm. Eq.~\eqref{eq:VarBound3Design} tells us that expectation values of operators with bounded Frobenius norm can be estimated to a good accuracy using a reasonable number of repetitions, regardless of what the input state. Importantly, for many observables (such as fidelities $\oper = \ket{\Psi}\bra{\Psi}$) this upper bound does not scale with system size.
	
	Measurement outcomes can be used to construct other properties of $\rho_S$ in addition to expectation values. Specifically, estimators for nonlinear functionals of $\rho_S$, such as R{\'e}nyi entropies, can also be obtained, as detailed in the following section. We note that bounds on the variance of such estimators proved in Ref.~\cite{Huang2020} also carry through to our protocol.
	
	\subsubsection{Blocking}
	
	To avoid a classical computational cost that scales exponentially in the size of $S$, it may be necessary to divide the system into $n$ blocks and entangle each with a set of decoupled ancillas $A_{j}$, $j = 1, \ldots, n$. In this case, the POVM operators have a tensor product structure $F_{m_1, \ldots, m_n} = \bigotimes_{j=1}^n F_{m_j}^{(j)}$, which precludes the formation of a state design over the entire Hilbert space. Instead, for generic choice of $U_j$, a state design within each block will be formed, and so the $k$th moments will have the form
	\begin{align}
		\mathcal{E}^{(k)} = \bigotimes_{j=1}^n \mathcal{E}_{\rm Haar}^{(k)}
		\label{eq:HaarBlock}
	\end{align}
	The corresponding channel $\mathcal{M}$ will have an analogous block structure $\mathcal{M} = \bigotimes_{j=1}^n (\text{id} + |\mathbbm{I}\rrangle \llangle \mathbbm{I}|)/(d_j+1)$, where $d_j$ is the Hilbert space dimension of block $j$.
	
	Obtaining a state-independent bound on the variance [analogous to \eqref{eq:VarBound3Design}] is more complicated in this case. If each block is a single qubit, $d_j = 2$, the problem becomes equivalent to bounding the variance of classical shadow tomography with random local Pauli measurements. This is because the uniform distribution over Pauli rotations forms a 3-design for a single qubit. Hence, the moments $k \leq 3$ of the POVM for local shadow tomography [Eq.~\eqref{eq:MomentRandom} are the same as the corresponding moments of the tomographic ensemble considered in our blocked protocol.
	
	In Ref.~\cite{Huang2020} it was shown that the variance can be upper bounded by $\Var \hat{\theta}_{\oper} \leq M^{-1} \|\oper\|_\infty 4^s$, where $s$ is the number of qubits for which $\oper$ acts non-trivially, and $\|C\|_\infty = \max \text{eig} \sqrt{C^\dagger C}$ is the spectral norm. Here we derive a bound that slightly improves on this, and can be generalized to any block Hilbert space dimension, including cases where each block is a different size. Writing the Hilbert space dimension of block $j$ as $d_j$, we have
	\begin{theorem}
		For any target state $\rho_S$ and observable $\oper$ that acts nontrivially on a subset of blocks $ T \subseteq \{1,\ldots, n\}$ [i.e.~$\oper = \mathbbm{I}_{\bar{T}} \otimes \oper_T$] with corresponding Hilbert space dimensions $d_j$, the variance of the estimator $\hat{\theta}_{\oper}$ constructed using a POVM for which the $k \leq 3$ moments of the tomographic ensemble are of the form \eqref{eq:HaarBlock} can be bounded as
		\begin{align}
			\Var \hat{\theta}_{\oper} &\leq \frac{1}{M} \|\oper\|_\infty^2 L(\{d_j\}_T) & \textrm{where } L(\{d_j\}_T) \leq \prod_{j \in T} \begin{dcases}
				2d_j & d_j \leq 4 \\
				\frac{3d_j^2}{d_j+2} & d_j > 4
			\end{dcases}
			\label{eq:BlockVar}
		\end{align}
		\label{thm:Block}
	\end{theorem}
	A full definition of function $L(\{d_j\}_T)$ is given in Eq.~\eqref{eq:LMin}; here we give an upper bound that takes a particularly simple form. When we set $d_j = 2$ for all $j$, the precise value of $L(\{2\}^{\times |T|})$ is $(3.7470\ldots)^{|T|}$, which gives a tighter bound than that proved for random Pauli measurements in Ref.~\cite{Huang2020}.\\
	
	\textit{Proof of Theorem \ref{thm:Block}.---}We begin by substituting \eqref{eq:HaarBlock} into our expression for the variance \eqref{eq:Variance3Design}, and separating the factors corresponding to $T$ and its complement $\bar{T}$
	\begin{align}
		\Var \hat{\theta}_\oper &= \frac{1}{M} \Tr\left[ \Big(\rho_S \otimes \big(\mathcal{M}_{\bar{T}}^{-1}[\mathbbm{I}_{\bar{T}}]\otimes \mathcal{M}_T^{-1}[\oper_T]\big)^{\otimes 2}\Big) \cdot \bigotimes_{j=1}^n \mathcal{E}^{(3)}_{{\rm Haar}, j} \right]
	\end{align}
	where $\mathcal{M}_T^{-1} = \bigotimes_{j \in T} \mathcal{M}_j^{-1}$ is the inverse map acting on blocks within $T$, and similarly for $\mathcal{M}_{\bar{T}}^{-1}$. Using the universal blocked form of the inverse map $\mathcal{M}_j^{-1}[C_j] = (d_j + 1)C_j - \Tr[C_j]\mathbbm{I}_j$ [which follows from \eqref{eq:HaarBlock}], we note that $\mathcal{M}_j^{-1}[\mathbbm{I}_j] = \mathbbm{I}_j$. We can therefore perform a partial trace over the blocks in $\bar{T}$, giving
	\begin{align}
		\Var \hat{\theta}_\oper &= \frac{1}{M}\Tr\left[ \Big(\rho_T \otimes \mathcal{M}_T^{-1}[\oper_T] \otimes \mathcal{M}_T^{-1}[\oper_T]\Big) \cdot \bigotimes_{j\in T} \mathcal{E}^{(3)}_{{\rm Haar}, j} \right]
	\end{align}
	where $\rho_T = \Tr_{\bar{T}} \rho_S$ is the reduced density matrix on $T$. Now we use the fact that $\mathcal{M}^{-1}$ is self-adjoint with respect to the Hilbert-Schmidt inner product, allowing us to move the inverse map onto the Haar moments
	\begin{align}
		\Var \hat{\theta}_\oper &= \frac{1}{M}\Tr\left[ \Big(\rho_T \otimes \oper_T \otimes \oper_T \Big) \cdot \bigotimes_{j\in T}\bigg[ \big(\text{id}_j \otimes \mathcal{M}^{-1}_j \otimes \mathcal{M}^{-1}_j\big)\Big[\mathcal{E}^{(3)}_{{\rm Haar}, j}\big] \bigg]\right]
		\label{eq:VarFactored}
	\end{align}
	We now aim to characterize the object inside the direct product in the above. Using the representation of the $k = 3$ moments of the Haar ensemble \eqref{eq:HaarLowMoments}, we can study the effect of the map $\mathcal{Q}_j \coloneqq (\text{id}_j \otimes \mathcal{M}^{-1}_j \otimes \mathcal{M}^{-1}_j)$ on each permutation operator $\pi_\sigma$ separately. Using cycle notation to denote the 6 elements of $\Sigma_3$ as $\{e, (1\,2), (2\,3), (1\,3), (1\,2\,3), (1\,3\,2)\}$, a straightforward calculation shows
	\begin{subequations}
		\begin{align}
			\mathcal{Q}_j[\pi_e] &= \pi_e \\
			\mathcal{Q}_j[\pi_{(1\,2)}] &= (d_j+1)^2 \pi_{(1\,2)} - (2d_j+1)\pi_e \\
			\mathcal{Q}_j[\pi_{(1\,3)}] &= (d_j+1) \pi_{(1\,3)} - \pi_e \\
			\mathcal{Q}_j[\pi_{(2\,3)}] &= (d_j+1) \pi_{(2\,3)} - \pi_e \\
			\mathcal{Q}_j[\pi_{(1\,2\,3)}] &= (d_j+1)^2 \pi_{(1\,2\,3)} - (d_j+1)[\pi_{(1\,3)} + \pi_{(2\,3)}] + \pi_e \\
			\mathcal{Q}_j[\pi_{(1\,3\,2)}] &= (d_j+1)^2 \pi_{(1\,3\,2)} - (d_j+1)[\pi_{(1\,3)} + \pi_{(2\,3)}] + \pi_e 
		\end{align}
	\end{subequations}
	After taking the required sum over permutations in $\mathcal{E}^{(3)}_j$, the direct product in \eqref{eq:VarFactored} becomes $\bigotimes_{j \in T} R_j$, where we define $R_j$ as
	\begin{align}
		R_j = (d_j+1)^2[\pi_{(1\,2)} + \pi_{(1\,2\,3)} + \pi_{(1\,3\,2)}] - (d_j+1)[\pi_e + \pi_{(1\,3)} + \pi_{(2\,3)}]
	\end{align}
	We now employ H{\"o}lder's inequality $\| X \cdot Y \|_1 \leq \|X\|_p \|Y\|_q$, where $\|X\|_p \coloneqq \Tr[|X|^p]^{1/p}$ is the $p$th Schatten norm of an arbitrary matrix $X$ with $|X| \coloneqq \sqrt{X^\dagger X}$, and $p, q \in [1, \infty]$ are chosen arbitrarily subject to the condition $1/p + 1/q = 1$. This gives an upper bound
	\begin{align}
		\Var \hat{\theta}_\oper &\leq \frac{1}{M}\| \rho_T \|_p \|\oper_T\|_p^2 \prod_{j\in T} \frac{\|R_j\|_q}{d_j(d+j+1)(d_j+2)} \nonumber\\
		&= \frac{1}{M}\|\oper_T\|_\infty^2 \prod_{j\in T} d_j^{2/p - 1} \frac{\|R_j\|_q}{(d_j+1)(d_j+2)}
		\label{eq:VarianceHolder}
	\end{align}
	where in the last step we use the inequality $\|X\|_p \leq n^{1/p} \|X\|_\infty$ for an $n \times n$ matrix $X$, as well as $\|\rho\|_p \leq 1$ for any valid density matrix and $p \leq 1$. To make use of this bound, we need to compute the Schatten norm, which is equal to the $p$-norm of the vector of singular values of $R_j$. By considering the action of $R_j$ on states of the form $\sum_{\sigma \in \Sigma_3} a_\sigma \ket{x_{\sigma(1)} x_{\sigma(2)} x_{\sigma(3)}}$, where $x_{1,2,3} \in \{1, \ldots, d_j\}$ label an orthonormal basis for the Hilbert space of block $j$, one can show that there are four distinct singular values of $R_j$, equal to 0, $3d_j(d_j+1)$, $(d_j+1)(d_j+2)$, and $2(d_j+1)(d_j+2)$, with respective degeneracies $2{d_j+1 \choose 3}$, ${d_j+2 \choose 3}$, ${d_j \choose 3}$, and $2{d_j+1 \choose 3}$.
	
	At this point we define the function $L(\{d_j\}_T)$ as the optimal value of the product in \eqref{eq:VarianceHolder}. Specifically,
	\begin{align}
		L(\{d_j\}_T) &= \min_{p \in [1,\infty]} \prod_{j \in T} \frac{d_j^{2/p - 1}}{(d_j+1)(d_j+2)}\left[ {d_j + 2 \choose 3} [3d_j(d_j+1)]^{p/(p-1)} \right. \nonumber\\ &+  {d_j \choose 3} [(d_j+1)(d_j+2)]^{p/(p-1)}  + \left. 2{d_j+1 \choose 3}[2(d_j+1)(d_j+2)]^{p/(p-1)} \right]^{(p-1)/p}
		\label{eq:LMin}
	\end{align}
	This expression is somewhat cumbersome, and so we calculate an upper bound of $ L(\{d_j\}_T)$ by setting $p = 1$, $q = \infty$ in which case we can replace $\|R_j\|_\infty$ with the maximum singular value, which is is $2(d_j+1)(d_j+2)$ for $d_j \leq 4$ and $3d_j(d_j+1)$ for $d_j \geq 4$. This is the result quoted in Theorem \ref{thm:Block}.
	
	We observe numerically that when $d_j \geq 4$ for all $j$, the minimum \eqref{eq:LMin} is obtained at $p = 1$, and so the inequality on the right hand side of \eqref{eq:BlockVar} becomes an equality. In the case of qubits $d_j = 2$, substituting the upper bound for $L(\{d_j\}_{T})$ given in \eqref{eq:BlockVar} reproduces the result $\Var \hat{\theta}_\oper \leq M^{-1} \|\oper\|_\infty^2 4^{|T|}$ found in Ref.~\cite{Huang2020}. In fact, by optimizing over $p$ we can find a tighter upper bound. Numerically we find that the optimal choice of $p$ is $1.1764\ldots$, which results in $\|R_j\|_p = 3.7470\ldots$.
	
	\subsection{Nonlinear functionals of $\rho_S$}
	
	For the majority of this paper, we have focussed on extraction of expectation values $\braket{\oper} = \Tr[\oper \rho_S]$, which are linear functionals of the system density matrix. Here, we describe how one can estimate nonlinear functionals of the form $\Tr[\oper_{(\ell)} \rho^{\otimes \ell}]$, where $\oper_{(\ell)}$ is an arbitrary operator acting on a $\ell$-fold replicated Hilbert space. Examples of such quantities are (exponentials of) R{\'e}nyi entropies $\exp[-(\ell-1)S^{(\ell)}] \coloneqq \Tr[\rho_S^{\ell}] \equiv \Tr[\pi_{(1\,2\,\ldots\, \ell)} \rho_S^{\otimes \ell}]$, as well as partially transposed moments $\Tr[(\rho_{AB}^{T_A})^\ell]$ (where $T_A$ denotes a partial transpose), which are used to construct entanglement negativities \cite{Vidal2002,Plenio2005,Elben2020}.
	
	The construction of nonlinear estimators here follows the same logic as those in Ref.~\cite{Huang2020}: First, for each repetition $r = 1, \ldots, M$ one constructs an unbiased estimator $\hat{\rho}_S^{(r)}$ of the full system density matrix, i.e.~$\mathbbm{E} \hat{\rho}_S^{(r)} = \rho_S$. In our case, the estimator can be constructed from a set of measurement outcomes using the optimal inverse map \eqref{eq:InvConstruct}
	\begin{align}
		|\hat{\rho}_S^{(r)}\rrangle = \mathcal{G}^* |m^{(r)})
		\label{eq:FullEstimator}
	\end{align}
	Using $\mathcal{G}^* \mathcal{F} = \text{id}$ and $\mathbbm{E}|m^{(r)}) = \mathcal{F} |\rho_S\rrangle$, we can show that the above is indeed an unbiased estimator of the system density matrix.
	
	Then, using these $M$ independent estimators, $U$-statistics can be used \cite{Ferguson2003} to construct an unbiased estimator of $\Tr[\oper_{(\ell)} \rho^{\otimes \ell}]$. To be specific, one considers all choices of $r_1 \neq r_2 \neq \cdots \neq r_\ell$, such that the estimators $\rho^{(r_1)}, \ldots, \rho^{(r_\ell)}$ are statistically independent. Each subset of runs can be used to construct an estimator, and taking an average over all these gives
	\begin{align}
		\hat{\theta}_{\oper_{(\ell)}} = \frac{1}{M(M-1)\cdots (M-\ell+1)} \sum_{r_1 \neq \cdots \neq r_\ell} \Tr\Big[ \oper_{(\ell)} \left(\hat{\rho}_S^{(r_1)} \otimes \cdots \otimes \hat{\rho}_S^{(r_\ell)}\right) \Big]
		\label{eq:EstNonlin}
	\end{align}
	The above is an unbiased estimator by virtue of the statistical independence of all $\hat{\rho}^{(r_i)}_S$ inside the trace.
	
	Determining the variance of \eqref{eq:EstNonlin} requires a somewhat more involved calculation than for linear estimators. Calculations of this kind have been performed in the context of conventional shadow tomography with randomized measurements \cite{Elben2020,McGinley2022}. Here we simplify matters by focusing on the $M \rightarrow \infty$ limit. Using the arguments of Ref.~\cite{Ferguson2003}, which apply generally to $U$-statistics of any kind, one can show that the variance takes an asymptotic form
	\begin{align}
		\Var \hat{\theta}_{\oper_{(\ell)}} =& \frac{\ell^2}{M} \Var \Tr\left[\oper_{(\ell)}^{\rm sym} \left( \hat{\rho}_S \otimes \rho_S^{\otimes (\ell -1)}\right) \right] + O\left(\frac{1}{M^2}\right)
	\end{align}
	where $\oper_{(\ell)}^{\rm sym} = (\ell !)^{-1}\sum_{\sigma \in \Sigma_\ell} \pi_\sigma \oper_{(\ell)} \pi_\sigma^\dagger$ contains only the parts of $\oper_{(\ell)}$ that act in a symmetric fashion on all replicas, and $\hat{\rho}_S$ is the estimator \eqref{eq:FullEstimator} for any choice of $r$. Using the explicit form of $\mathcal{G}^*$ in \eqref{eq:InvConstruct}, we have
	\begin{align}
		\Var \hat{\theta}_{\oper_{(\ell)}} =& \frac{\ell^2}{M} \sum_m p_m \Tr\left[\oper_{(\ell)}^{\rm sym} \left( \mathcal{G}_0[m] \otimes \rho_S^{\otimes (\ell -1)}\right) \right]^2 + O\left(\frac{1}{M^2}\right) \nonumber\\
		=& \frac{\ell^2}{M} \Tr\left[\Big(\rho_S \otimes \mathcal{M}^{-1}[\oper_1]^{\otimes 2}\Big) \mathcal{E}^{(3)} \right] + O\left(\frac{1}{M^2}\right)
	\end{align}
	where $\mathcal{O}_1 \coloneqq \Tr_{2,\ldots, \ell}[\oper_{(\ell)}^{\rm sym} \cdot (\mathbbm{I}_1 \otimes \rho_S^{\otimes (\ell-1)})]$ contains the partial trace over $(\ell-1)$ replicas, and $\mathcal{E}^{(3)}$ is the third moment of the tomographic ensemble \eqref{eq:Ensemble}. Again, this variance depends only on the $k \leq 3$ moments of the tomographic ensemble, and so when the ensemble forms a 3-design (which generically occurs in our protocol without conservation laws), the variance will be the same as the corresponding estimators in conventional shadow tomography.
	
\end{onecolumngrid}


\end{document}